\documentclass[amsmath,amssymb,showpacs,twocolumn]{revtex4}
\usepackage{epsfig,amsmath}

\begin{document}

\author{M.~F.~Gelin}
\affiliation{Department of Chemistry, Technische Universit\"at M\"unchen, D-85747 Garching, Germany}

\author{I.~V.~Bondarev}
\affiliation{Department of Mathematics and Physics, North Carolina Central University, Durham,
NC 27707, USA}

\title{One-dimensional transport in hybrid metal-semiconductor nanotube systems}
\begin{abstract}
We develop an electron transport theory for the hybrid system of a semiconducting carbon nano\-tube that encapsulates a one-atom-thick metallic wire. The theory predicts Fano resonances in electron transport through the system, whereby the interaction of electrons on the wire with nano\-tube plasmon generated near-fields blocks some of the wire transmission channels to open up the new coherent plasmon-mediated channel in the nanotube forbidden gap outside the wire transmission band. Such a channel makes the entire hybrid system transparent in the energy domain where neither wire, nor nanotube is indivudually transparent. This effect can be used to manipulate by the electron charge transfer in hybrid nanodevices built on metal-semiconductor nanotube systems.
\end{abstract}
\maketitle

\section{Introduction}

Over the last decade, electron transport studies in quasi-one-dimensional (1D) nano\-structures have resulted in important discoveries, such as conductance quantization and oscillatory length dependence, molecular rectification, negative differential resistance, hysteresis behavior, etc.\cite{agrait03,datta04,nitzan03,Zimbovskaya11} At present, peculiar mechanisms of electron transport are fairly well understood for pristine atomic wires (AWs)~\cite{agrait03}, carbon nanotubes (CNs) and some CN-based components~\cite{Zimbovskaya11,Roche07}. Carbon nanotubes --- graphene sheets rolled-up into cylinders of one to a few nanometers in diameter and up to hundreds of microns in length~\cite{Saito} --- have been successfully integrated into miniaturized electronic, electromechanical, chemical devices and into nanocomposite materials~\cite{Dresselhaus01,Baughman13}, and have found a variety of applications in optoelectronics~\cite{Avouris08,McEuen09,Mueller10,Hertel10,Malic11,Kamat07,Strano11,Kono12,ChemPhysSI}.

\begin{figure}
\epsfxsize=8.3cm\centering{\epsfbox{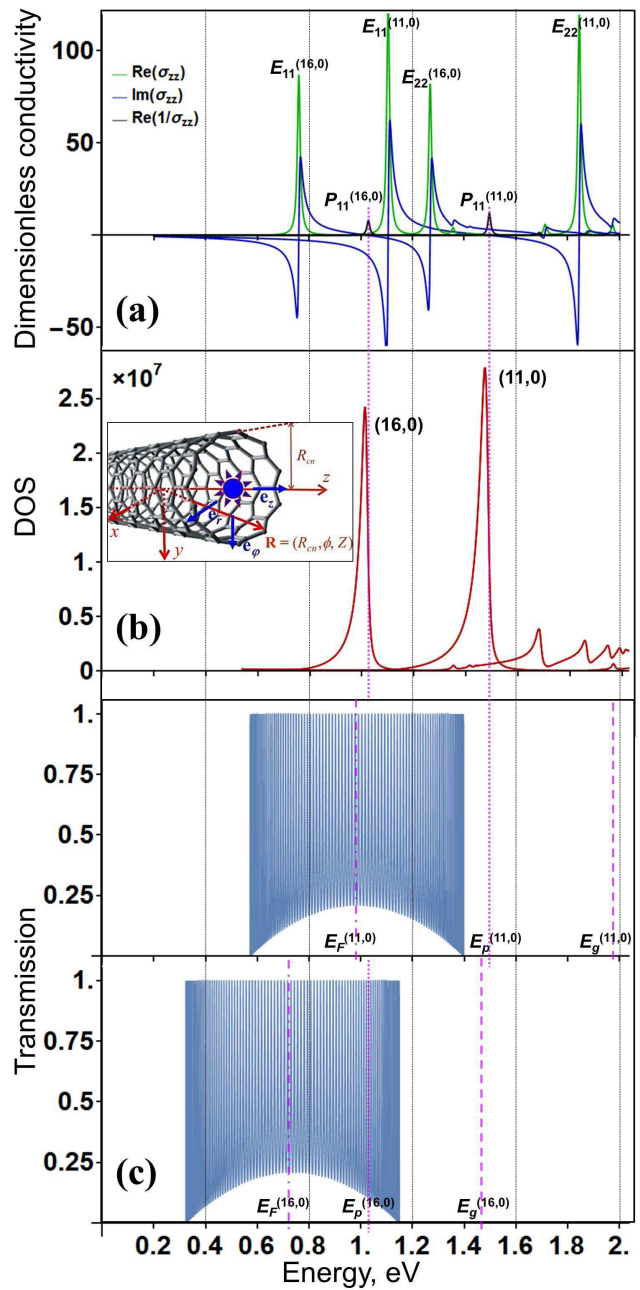}}\caption{(Color online) (a)~Dimensionless (normalized by $e^2/2\pi\hbar$) surface axial conductivities $\sigma_{zz}$ for the semiconducting (11,0) and (16,0) CNs. Peaks of $\mbox{Re}\,\sigma_{zz}$ and $\mbox{Re}(1/\sigma_{zz})$ represent excitons ($E_{11}$, $E_{22}$) and interband plasmons ($P_{11}$), respectively. (b)~Photonic DOS for non-radiative spontaneous decay with CN plasmon excitation for a two-level dipole emitter on the symmetry axis (inset) of the (11,0) and (16,0) CN. (c)~Electron transmission band for the \emph{free} AW of 100 sodium atoms with energy counted from the bottom of the fundamental band gap $E_g$ of the (11,0) and (16,0) CNs (top and bottom, respectively); also shown are the AW Fermi energies $E_F$ and CN first interband plasmon energies $E_p$. Conductivities are obtained using the ($\mathbf{k}\cdot\mathbf{p}$)-scheme by Ando~\cite{Ando}. DOS functions are calculated as described by Bondarev and Lambin in Refs.~\cite{Bondarev05,BondarevNova06}. Transmission is plotted within the nearest-neighbor hopping model as discussed by Mujica \emph{et al.} and by Gelin and Kosov in Refs.~\cite{Muj94a,Muj94b,GZK}. See Sec.~\ref{sec5} for more details.}\label{fig1}
\end{figure}

Enormous potential of carbon nanotubes as building blocks for designing optoelectronic nanodevices stems from their extraordinary thermo\-mecha\-nical stability combined with unique physical properties that originate from quasi-one-dimensionality to give rise to a peculiar quasi-1D band structure featuring intrinsic, spatially confined, collective electronic excitations such as excitons and plasmons~\cite{Ando,Dresselhaus07,Pichler98,Marin03,Bondarev09,BondarevNova11}. Due to the circumferential quantization of the longitudinal electron motion, real axial (along the CN axis) optical conductivities of single wall CNs consist of series of peaks $E_{11}, E_{22}, ...$, representing the first, second, etc. excitons, respectively [see Fig.~\ref{fig1}~(a)]. Imaginary conductivities are linked with the real ones by the Kramers-Kronig relation, and so real \emph{inverse} conductivities show the resonances $P_{11}, P_{22}, ...$ next to their excitonic counterparts. These are weakly dispersive, low-energy ($\sim\!1\!-\!2$~eV) inter-band plasmon modes. They were observed experimentally quite a while ago~\cite{Pichler98}, and were theoretically demonstrated quite recently to play the key role in a variety of new surface electromagnetic (EM) phenomena with CNs~\cite{Bondarev09,Popescu11,BondarevNova11,Bondarev12,Bondarev12pss,Bondarev04,Bondarev05,Bondarev06,BondarevNova06,Bondarev07,GelinBondarev13,Woods13,Bondarev14,GelinBondarev14,Bondarev15OE}, such as exciton-plasmon coupling~\cite{Bondarev09,BondarevNova11}, plasmon generation by excitons~\cite{Bondarev12,Bondarev12pss}, exciton Bose-Einstein condensation in individual single wall CNs~\cite{Bondarev14}, Casimir attraction in double wall CNs~\cite{Popescu11,BondarevNova11,Woods13}, van der Waals coupling~\cite{Bondarev05,BondarevNova06}, spontaneous emission~\cite{BondarevNova06,Bondarev04}, resonance absorption~\cite{Bondarev06}, scattering~\cite{Bondarev15OE}, bipartite entanglement in hybrid systems of extrinsic atoms/ions doped into CNs~\cite{Bondarev07,GelinBondarev13,GelinBondarev14}, to mention a few, --- all of relevance to conceptually new tunable optoelectronic device applications with CNs~\cite{Bondarev07jem,Bondarev10JCTN,Bondarev14Dekker}.

Apparently, further progress in CN optoelectronics can be expected from the exploration of complex hybrid CN structures~\cite{Tasis10,Kharlamova13}, particularly, CNs doped with extrinsic species such as molecules~\cite{Duclaux02,Wim,Cambre,Iwasa03}, semiconductor quantum dots~\cite{Giersig,Krauss,Kamat07}, atoms and ions~\cite{Duclaux02,Kawazoe03,Kawazoe04,Barrera11}, and also AW-encapsulating CNs~\cite{Shinohara09,Russkie09,Russkie,Crampin11,Pascard94,Morales99,Schneider03,Walton98,Chang96,Pascard96,Iijima07,Lijima08,Shinohara08,Sumpter09,Dresselhaus08,Shinohara12,Hatakeyama07}. This research direction is being pursued by a number of groups worldwide. Carbon nanotubes of different diameters are synthesized to host various metallic AWs, including Cr~\cite{Pascard94}, Fe~\cite{Morales99}, Co~\cite{Schneider03}, Ni~\cite{Walton98}, Cu~\cite{Chang96}, Ge~\cite{Pascard96}, I~\cite{Iijima07}, La~\cite{Lijima08}, Gd~\cite{Shinohara08}, Mo~\cite{Sumpter09,Dresselhaus08}, Eu~\cite{Shinohara12}, and Cs~\cite{Hatakeyama07}. Quantum chemistry simulations are performed for the electronic structure of AW-encapsulating CNs~\cite{Lu03,Seifert07,Kim10,Belucci13,Lee08,Jo09,Gao11,Vega11,Zhu09}, supplemented with electron transmission calculations for CNs encapsulating Gd/Eu~\cite{Gao11}, Mo~\cite{Vega11} and Au-V(Cr)~\cite{Zhu09} AWs.~However, the inter-play between the intrinsic 1D conductance of an atomic wire and CN plasmon mediated near-fields is still far from being well understood, calling for deeper theoretical insight into electron transport peculiarities in these complex hybrid quasi-1D quantum systems.

Encapsulating metallic wires of just one atom thick into a single wall CN, metallic or semiconducting, is known to drastically alter the transport properties of the compound hybrid system. The whole body of available data shows that transport peculiarities in hybrid CN--AW systems cannot be explained by a mere addition of the properties of pristine CNs and AWs, suggesting a crucial role of the CN--AW interactions with a variety of associated quantum interference effects, including opening extra (collective) transport channels~\cite{Gao11,Vega11,Zhu09,Shinohara12}. For example, metallic single wall CNs encapsulating Eu atomic wires are experimentally demonstrated to have extra conduction channels to supplement an overall "Tomonaga-Luttinger liquid"-like transport behavior~\cite{Shinohara12}, and with both CNs and AWs being non-metals their compound CN--AW hybrids are theoretically predicted to be metallic for CNs of appropriately chosen diameters~\cite{Vega11}.

Pristine metallic single wall CNs are excellent conductors by themselves~\cite{Roche07}, and so, theoretically, there is no surprise about extra channels in the CN transport that emerge upon encapsulating metal AWs into them. From theoretical perspective, more interesting is a hybrid system of metal-atomic-wire encapsulating \emph{semiconducting} CNs. In these systems, at low bias voltages not exceeding the CN fundamental band-gap, the CN itself does not have any intrinsic open channels to conduct electrons. Hence, there is no electron exchange between the CN and the metallic AW encapsulated into it, with the transport being totally dominated by the AW alone, while at the same time being affected by local quasi-static fields of nanotube's collective interband plasmon excitations.

As an example, Fig.~\ref{fig1} shows same scale energy dependences for characteristic parameters to represent semiconducting CNs, their near-field interaction with encapsulated atomic types species, and one-dimensional (1D) metallic atomic wires, respectively.~Panel~(a) shows the dynamical axial conductivities for the semiconducting (11,0) and (16,0) CNs. Shown are the real conductivities, the imaginary conductivities, and the real inverse conductivities of relevance to the electron energy-loss spectroscopy response function used in studies of collective plasmon excitations in solids~\cite{Bondarev09,BondarevNova11,Marin03,Pichler98,Toyozawa}. Peaks of the real conductivities and those of the real inverse conductivities represent excitons ($E_{11}$, $E_{22}$) and interband plasmons ($P_{11}$), respectively~\cite{Bondarev09,BondarevNova11,Bondarev12,Bondarev12pss,Bondarev14,Popescu11,Woods13}. Panel~(b) shows local densities of photonic states (DOS) for the non-radiative spontaneous decay (relative to vacuum) of a two-level dipole emitter placed on the symmetry axis (inset) inside the (11,0) and (16,0) CNs. Panel~(c) shows the electron transmission band for the free 1D metallic AW of 100 sodium atoms with energy counted from the bottom of the fundamental band gap of the (11,0) and (16,0) CNs (top and bottom, respectively). All AW transport channels are seen to be inside of the CN forbidden gap. However, AWs encapsulated into CNs will experience near-field EM coupling to the CN interband plasmon modes represented by the large DOS resonances in panel~(b). Inter\-band plasmons are \emph{standing} charge density waves due to the periodic opposite-phase displacements of the electron shells with respect to the ion cores in the neighboring elementary cells on the CN surface. Such periodic displacements induce local coherent oscillating electric fields of zero mean, but nonzero mean-square magnitude, concentrated near the surface across the diameter throughout the length of the CN~\cite{Bondarev12,Bondarev12pss}. The electric dipole interaction of atoms on the wire with these near-fields will largely affect the AW transmission properties as well as the properties of the entire hybrid system.

The coupling strength of an individual atom to the plasmon-induced near-fields inside the nanotube can be estimated as follows. Modeled by a two-level system, an atom (ion) interacts with the CN medium-assisted fields via an electric dipole transition $d_{z}\!=\!\langle u|\hat{d}_z|l\rangle$ between its lower $|l\rangle$ and upper $|u\rangle$ states~\cite{Bondarev05,BondarevNova06}, with the CN symmetry axis set to be the $z$-quan\-ti\-za\-tion axis [Fig.~\ref{fig1}(b), inset]. Transverse dipole orientations can be neglected in view of the strong transverse depolarization in individual CNs~\cite{Tasaki98,Li01,Bondarev02,Marin03}. The atom--CN electric dipole coupling constant is then given by $\hbar g=(2\pi d_z^{\,2}\hbar\tilde{\omega}_{A}/\tilde{V})^{1/2}\!$~\cite{Andreani,Bondarev06}, where $\tilde{\omega}_{A}$ is the effective transition frequency \emph{in resonance} with a local CN-medium-assisted field mode.~The effective mode volume $\tilde{V}\!\!\sim\!\pi R_{cn}^{2}(\tilde{\lambda}_{A}/2)$ for the CN of radius $R_{cn}$. Evaluating $d_z\!\sim\!er_A\!\sim\!e(e^{2}/\hbar\tilde{\omega}_{A})$, where $r_A$ is the linear size of the atom (estimate valid for quantum systems with Coulomb interaction~\cite{Davydov}), and introducing the fine structure constant $\alpha\!=\!e^2/\hbar c\!=\!1/137$, one arrives at $\hbar g\!=\!(2\alpha^3/\pi)^{1/2}(\hbar c/R_{cn})$, to give $\hbar g\!\sim\!0.2$~eV for CNs with diameters $\sim\!1$~nm.~Comparing this to the "cavity" linewidth $\hbar\gamma_{c}(\tilde{\omega}_{A})\!=\!6\pi\hbar c^{3}/\tilde{\omega}_{A}^{2}\xi(\tilde{\omega}_{A})\tilde{V}$ [function $\xi$ represents the local photonic DOS at the atomic location, also called the Purcell factor~\cite{Andreani}; shown in Fig.~\ref{fig1}(b)], one has the ratio $g/\gamma_{c}\!\sim\!10\!\gg\!1$ for the 1~nm-diameter CNs in the optical spectral range. This is the strong atom-field coupling condition characterized by the rearrangement ("dressing") of atomic levels and formation of atomic quasi-1D cavity polaritons~\cite{Bondarev05,BondarevNova06} featuring strongly non-exponential (oscillatory) spontaneous decay dynamics~\cite{BondarevNova06,Bondarev04} and Rabi splitting of the optical absorption line profile~\cite{Bondarev06}. However, one can hardly expect such a strong resonance coupling for identical atoms aligned on a wire inside the CN as the $|l\rangle$ and $|u\rangle$ transition states are here a subset of the one-dimensional electronic band states, and so are not as clearly defined as those for an individual atom, to most likely result in a weaker \emph{non-resonance} CN--AW coupling strength.

Here we study the inter-play between the intrinsic 1D conductance of metallic AWs and CN mediated near-field effects for \emph{semiconducting} single wall CNs that encapsulate atomic wires of finite length. We use matrix Green's functions formalism to develop an electron transfer theory for such a complex hybrid quasi-1D CN system. The theory predicts Fano resonances in electron transmission through the system. That is the AW--CN near-field interaction blocks some of the pristine AW transmission band channels to open up new coherent channels in the CN forbidden gap outside the pristine AW transmission band. This makes the entire hybrid system transparent in the energy domain where neither of the individual pristine constituents, neither AW nor CN, is transparent. The effect can be used to control electron charge transfer in semiconducting CN based devices for nanoscale energy conversion, separation and storage~\cite{Cola15,Zwiller14,Kono13,deHeer98}.

The paper is structured as follows. Section~\ref{sec2} formulates the theoretical model for CN-mediated AW transmission. Section~\ref{sec3} presents analytical expressions derived for the transmission coefficient. The expressions obtained are analyzed qualitatively in Section~\ref{sec4}, and then numerically in Section~\ref{sec5}. Section~\ref{sec6} discusses the model approach and the approximations used. A brief summary of the work is given in Section~\ref{sec7}. Appendix~\ref{ApA} derives mathematical expressions presented in Section~\ref{sec3}.

\section{The model\label{sec2}}

This Section formulates the model for the plasmon-mediated electron transport through the hybrid metal AW encapsulating single wall semiconductor CN system. The detailed physical justification for the model and its theoretical interpretation are discussed in Section~\ref{sec6}.

\subsection{The Hamiltonian of the hybrid CN--AW system}

The quantum system under consideration is a metallic AW encapsulated into a single-wall semiconducting CN. The system is assumed to be attached to two electrodes (leads) kept at a certain bias voltage. The quantum mechanical observable of interest is the steady-state electron current through the CN--AW system.

We write the total Hamiltonian of the entire system as the sum of the Hamiltonians for the AW, the CN, and their interaction as follows
\begin{equation}
\hat{H}=\hat{H}_{\text{AW}}+\hat{H}_{\text{CN}}+\hat{H}_{\text{int}}.
\label{Htot}
\end{equation}

To describe the AW of $N$ atoms (lattice sites) in length, we adopt the standard second quantized tight-binding model Hamiltonian with the nearest neighbor electron hopping rate $V$ and the electron on-site energy $E_{0}$,
\begin{equation}
\hat{H}_{\text{AW}}=E_0\sum_{k=1}^N B^{\dagger}_kB_k+V\sum_{k=1}^{N-1}\!\left(B^{\dagger}_k B_{k+1}+B^{\dagger}_{k+1}B_k\right).
\label{HAW}
\end{equation}
Here, the operators $B^{\dagger}_k$ and $B_k$ create and annihilate, respectively, the electronic excitations on site $k$ of the AW (see, e.g., Refs.~\cite{Muj94b,Muj94a,GZK}). They obey the Pauli commutation rules $[B_{k},B_{n}^{\dagger}]=\delta_{kn}(1-2B_{k}^{\dagger}B_{n})$.

We assume the bias voltage not to exceed the fundamental bandgap $E_g\,[\,=\!E_g^{(11)}]$ of the semiconducting CN as it shows in Fig.~\ref{fig1}~(a)--(c). Then, regardless of whether the AW transmission band is narrower [as in Fig.~\ref{fig1}~(c)] or broader than $E_g$, the electrons are transported in between the leads at energies $E_F\!\lesssim\!E\!<\!E_g$ inside the forbidden gap of the CN. As this takes place, the CN does disturb the transport due to the quasi-static near-fields of weakly dispersive, low-energy ($\sim\!1\!-2$~eV) collective interband plasmon modes represented by the local photonic DOS resonances as shown in Fig.~\ref{fig1}~(b).

The most general second quantized Hamiltonian of the CN can then be written as follows (Ref.~\cite{Bondarev09})
\begin{equation}
\hat{H}_{\text{CN}}\!=\!\sum_\mathbf{n}\int_0^\infty\!\!\!d\omega\,\hbar\omega\hat{f}^\dag(\mathbf{n},\omega)\hat{f}(\mathbf{n},\omega).
\label{HCN}
\end{equation}
In this equation, the scalar bosonic field operators $\hat{f}^\dag(\mathbf{n},\omega)$ and $\hat{f}(\mathbf{n},\omega)$ create and annihilate, respectively, the surface EM excitation of the frequency $\omega$ at an arbitrary point $\mathbf{n}\!=\!\mathbf{R}_n\!=\!\{R_{CN},\varphi_n,z_n\}$ that represents the position of a carbon atom [nanotube lattice site --- Fig.~\ref{fig1}~(b), inset] on the surface of the CN of radius $R_{CN}$, $[\hat{f}(\mathbf{n},\omega),\hat{f}^\dag(\mathbf{n}^\prime,\omega^\prime)]\!=\!\delta_{\mathbf{n}\mathbf{n}^\prime}\delta(\omega-\omega^\prime)$. Summation is taken over all the carbon atoms on the entire CN surface. Since only one plasmon resonance is located inside $E_g$, Eq.~(\ref{HCN}) can further be simplified to take the form
\begin{equation}
\hat{H}_{\text{CN}}=E_{p}\,\hat{f}^{\dagger}\hat{f}.
\label{HCNfin}
\end{equation}
Here, the operators $\hat{f}^{\dagger}$ and $\hat{f}$ ($[\hat{f},\hat{f}^{\dagger}]\!=\!1$) create and annihilate \emph{collective} interband plasmon excitations of energy $E_{p}$ [Fig.~\ref{fig1}~(a),(b)] that are delocalized all over the CN surface in accordance with the correspondence relation
\begin{equation}
\sum_\mathbf{n}\hat{f}^\dag(\mathbf{n},\omega)\hat{f}(\mathbf{n},\omega)=\hat{f}^{\dagger}\hat{f}\,\delta(\omega-E_p/\hbar).
\label{corresp}
\end{equation}
The CN--AW interaction can then be written in the form
\begin{equation}
\hat{H}_{\text{int}}=\sum_{k=1}^{N}\mu_{k}\left(B_{k}\hat{f}^{\dagger}+B_{k}^{\dagger}\hat{f}\right),
\label{Hint}
\end{equation}
where $\mu_{k}$ is the AW--CN dipole coupling constant for site $k$ of the AW. We assume it to be the same for all of the AW sites, that is
\begin{equation}
\mu_k\!=\!\mu\lesssim\hbar g=\sqrt{\frac{2\pi d_z^{\,2}\hbar\tilde{\omega}_{A}}{\tilde{V}}}\approx\sqrt{\frac{2\alpha^3}{\pi}}\frac{\hbar c}{R_{cn}}
\label{mu}
\end{equation}
in what follows, as discussed in the introduction above.

The Hamiltonian in Eqs.~(\ref{Htot})--(\ref{mu}) belongs to the well-known family of Fano-Anderson Hamiltonians~\cite{Mahan,Fano}. However, it describes a physical picture opposite to that normally refereed to as the Fano--Anderson model. The latter deals with a \emph{bound} (e.g., localized on a defect) electron state inside (or outside) of the continuum of scattering (band, or free) electron states~\cite{Mahan}. In our case, the band electron states represented by the AW Hamiltonian~(\ref{HAW}), interact with the CN collective interband plasmon excitations described by the Hamiltonian~(\ref{HCNfin}). These are standing charge density waves due to the periodic opposite-phase displacements of the electron shells with respect to the ion cores in the neighboring elementary cells on the CN surface~\cite{Bondarev12,Bondarev12pss}. They are \emph{extended} coherently all over the entire surface of the CN as described by Eqs.~(\ref{HCN}) and (\ref{corresp}). The main feature of the standard Fano--Anderson model is still there though, offering two different electron transmission paths. They are: (i)~the direct transfer through the AW, and (ii)~the transfer mediated by quasi-static near-fields due to the CN collective interband plasmon excitations.~Thus, the model we present here is a non-trivial extension of the Fano-Anderson model to cover coherently delocalized electron states, such as collective plasmon excitations, in addition to localized (defect-type) states studied originally.

\subsection{The matrix representation of the operators}

Using the relevant single-quantum Hilbert space basis of $N\!+\!1$ basis vectors as follows
\[
\left\{B_{k}^{\dagger}\,|0\rangle\right\}_{\!k\,=1,\cdots,N}\,,\,\hat{f}^{\dagger}|0\rangle\,,
\]
with $|0\rangle$ being the vacuum state of the entire system, one can convert the total Hamiltonian~(\ref{Htot}) into the matrix representation as follows
\begin{equation}
\mathbf{H}=\left[\begin{array}{ccccccc}
E_{0} & V & 0 & \ldots & 0 & 0 & \mu\\
V & E_{0} & V & \ldots & 0 & 0 & \mu\\
0 & V & E_{0} & \ldots & 0 & 0 & \mu\\
\vdots & \vdots & \vdots & \ddots & \vdots & \vdots & \vdots\\
0 & 0 & 0 & \ldots & E_{0} & V & \mu\\
0 & 0 & 0 & \ldots & V & E_{0} & \mu\\
\mu & \mu & \mu & \ldots & \mu & \mu & E_{p}
\end{array}\right].
\label{Htotmatrix}
\end{equation}
Here, rows and columns $1,\dots,N$ refer to the AW, enumerating its sites, and the $(N\!+\!1)$-st row/column refers to the CN interband plasmon mode. Similar matrix representations can be written for any other operator of relevance to the problem.

Matrix~(\ref{Htotmatrix}) represents the tight-binding Hamiltonian for the entire hybrid CN--AW system. Its diagonal matrix elements $H_{11},\dots,H_{NN}$ are the site energies of the AW sites that incoming electrons hop through; its off-diagonal elements specify the rates at which the hopping occurs through the AW, and its eigen energies determine the resonances of the electron transmission through the CN--AW system. The CN--AW coupling term in the $(N\!+\!1)$-st row/column modifies resonance transmission energies and hopping pathways. To explore the role of this latter ingredient of the model is the goal of this work.

\subsection{The transmission coefficient}\label{ET}

To calculate the transmission coefficient through the metal AW encapsulated into a semiconducting CN, we follow the Landauer formalism~\cite{agrait03,datta04,nitzan03,Zimbovskaya11}, in which the presence of the leads is accounted for by the electron self-energy operator
\begin{equation}
\mathbf{\Sigma}(E)=\mathbf{\Lambda}(E)-i\mathbf{\Delta}(E),
\label{selfen}
\end{equation}
Here, the real and imaginary part represents the shift and broadening, respectively, of the \emph{terminal} electron energy level $E$ due to the coupling between the wire and the leads. This operator enters the Green's function of our system as follows~\cite{Muj94a,Muj94b,GZK,GK_as}
\begin{equation}
\mathbf{G}(E)=[E-\mathbf{H}-\mathbf{\Sigma}(E)]^{-1}.
\label{Trdef}
\end{equation}
With the left and right leads attached to the 1-st and $N$-th AW site, respectively, the transmission probability is proportional to the Green's function matrix element $G_{1N}$, according to the general scattering matrix formalism as applied to atomic and molecular wires~\cite{Muj94a}. Then, the electron transmission coefficient $T(E)$ is given by
\begin{equation}
T(E)=4\Delta^{2}(E)\left|G_{1N}(E)\right|^{2}\!.\label{T}
\end{equation}
Here, $\Delta$ is the imaginary part of the self-energy function $\Sigma(E)$ defined in terms of the only two non-zero matrix elements of the self-energy operator (\ref{selfen}) as follows
\begin{equation}
\Sigma(E)=\Lambda(E)-i\Delta(E)\equiv\Sigma_{11}(E)\equiv\Sigma_{NN}(E),
\label{selfenfunc}
\end{equation}
with the left and right leads assumed to be identical.

The transmission coefficient (\ref{T}) is the probability for an electron to be transferred between the leads at a constant energy $E$, that is $0\!\le T\!\le\!1$. The conductance of the CN--AW system in the linear regime, whereby voltages are small and temperatures are low, is given by the Landauer formula (in units of $e^{2}/2\pi\hbar$)
\begin{equation}
\mbox{g}=T(\epsilon_{F}),
\label{g}
\end{equation}
where $\epsilon_{F}$ is the Fermi energy of the leads~\cite{Muj94a}.~This quantity only depends on the electronic structure of the wire and the leads, and does not depend on the field applied.

\section{The transmission of electrons: exact analytical solution \label{sec3}}

To evaluate the matrix element $G_{1N}$ of the Green's function~(\ref{Trdef}) in Eq.~(\ref{T}), we introduce the matrix
\[
\mathbf{h}\equiv E-\mathbf{H}-\mathbf{\Sigma}(E),
\]
and follow the rules of calculating the matrix elements of its inverse. One obtains
\begin{equation}
G_{1N}(E)=\frac{\mathbf{h}_{N1}}{\det(\mathbf{h})},
\label{G1Ndef}
\end{equation}
with $\mathbf{h}_{N1}$ representing the $N1$ cofactor of the matrix $\mathbf{h}$. Next, we use the result of the matrix Green's function partitioning technique developed in Ref.~\cite{Muj94a}, whereby
\begin{equation}
\det(\mathbf{h})=\Sigma^2 D_{N-2} + 2\,\Sigma D_{N-1} + D_N
\label{deth}
\end{equation}
with $\Sigma$ given by Eq.~(\ref{selfenfunc}), and
\begin{equation}
\mathbf{h}_{N1}=(-1)^{N+1}S_{N-1},
\label{hN1}
\end{equation}
with $D_N$ and $S_{N-1}$ being the determinant and the $N1$ minor of the matrix $\mathbf{H}\!-E$, respectively (both determined by the AW lattice site number $N$; subscripts to indicate that the latter is a polynomial of degree one less than the former). The matrix $\mathbf{H}$ is given by Eq.~(\ref{Htotmatrix}).

Next, if we make a standard assumption that the identical leads are made of a broadband metal with the half-filled conduction band, as discussed in Ref.~\cite{Muj94b}, then the real part $\Lambda$ of the self-energy function~(\ref{selfenfunc}) cancels out. The imaginary part $\Delta$ takes the energy independent form $\Delta\!=\!V_S^2/\gamma$, with $V_S\!\equiv\!V_1\!=\!V_N$ representing the lead--wire terminal chemisorption coupling constant and $\gamma$ being the lead metal half-bandwidth. In view of Eqs.~(\ref{G1Ndef})--(\ref{hN1}), the transmission coefficient~(\ref{T}) then takes the form
\begin{equation}
T(E)=4\Delta^{2}\left|\frac{S_{N-1}}{\Delta^2 D_{N-2} + 2i \Delta D_{N-1} - D_N}\right|^{\,2}.
\label{TEfin1}
\end{equation}
Here, the quantities $D_N$ and $S_N$ are functions of the AW lattice site number $N$, which are given by the analytical expressions as follows (see Appendix~\ref{ApA} for the derivation)
\begin{equation}
D_N=\varepsilon_p\,d_N\,+
\label{DN}
\end{equation}
\[
\frac{\mu^2}{\varepsilon_0+2V}\!\left\{\!-N d_N+\frac{2V}{\varepsilon_0+2V}\!\left[(-1)^N V^N\!-Vd_{N-1}\!-d_N\right]\!\right\}\!,
\]
\begin{equation}
S_N=\varepsilon_p V^N\!+\frac{\mu^2}{\varepsilon_0+2V}\!\left[-(N+1)V^N\!+(-1)^N d_N\right]\!,
\label{SN}
\end{equation}
where
\begin{equation}
\varepsilon_0=E_0-E, \,\,\, \varepsilon_p=E_p-E\,,
\label{eps0}
\end{equation}
and
\begin{equation}
d_{N}=\frac{\lambda_1^{N+1}-\lambda_2^{N+1}}{\lambda_1-\lambda_2}
\label{dN}
\end{equation}
with
\begin{equation}
\lambda_{1,2}=\frac{\varepsilon_0 \pm \sqrt{\varepsilon_0^2-4V^2}}{2}\,.
\label{lambda12}
\end{equation}

\section {Qualitative analysis \label{sec4}}

The transmission coefficient given by Eqs.~(\ref{TEfin1})--(\ref{lambda12})~is the key quantity to describe the electron transfer through the atomic wire encapsulated into a carbon nanotube. There are two parameters to control the AW--CN coupling there.~They are the atom-plasmon coupling constant $\mu$ and the plasmon energy detuning $\varepsilon_{p\,}$ in Eqs.~(\ref{mu}) and (\ref{eps0}), respectively. From Eqs.~(\ref{DN}) and (\ref{SN}), we see that it~is the ratio $\mu^2/\varepsilon_p=\mu^2/(E_p-E)$ that determines the $T(E)$ energy dependence. [The conductance (\ref{g}) is determined by $\mu^2/(E_p-\epsilon_F)$, accordingly.] In view of this, increasing $\mu$ affects $T(E)$ the same way as decreasing $\varepsilon_p$, and vice versa. Therefore, we restrict ourselves to the analysis of the $T(E)$ behavior versus $\mu$ in what follows.

\subsection {Pristine AWs \label{PrAW}}

The pristine AW case follows from the general equation~(\ref{TEfin1}) if one substitutes $\mu\!=\!0$ there. Then, we recover the known quantum wire transmission formula~\cite{Muj94a,GZK}
\begin{equation}
T(E)=4\Delta^{2}\left|\frac{V^{N-1}}{\Delta^2 d_{N-2} + 2i \Delta d_{N-1} - d_N}\right|^{2}.
\label{TEpristine}
\end{equation}
This shows that there are two possible, qualitatively different electron transport regimes there for pristine AWs, depending on whether $|\varepsilon_0/2V|\!<\!1$, or $|\varepsilon_0/2V|\!>\!1$.

In the case where $|\varepsilon_0/2V|\!<\!1$, the roots $\lambda_{1,2}$ of Eq.~(\ref{lambda12}) are complex, yielding $d_N$ in Eq.~(\ref{dN}) of the form
\begin{equation}
d_{N}=\frac{\sin[(N+1)\phi]}{\sin\phi}\,V^N
\label{dNpristine}
\end{equation}
with $\phi$ given by the roots of the equation $\cos\phi\!=\!\varepsilon_0/2V$. In this case Eq.~(\ref{TEpristine}) takes the form
\[
T(E)\!=\!\left|\frac{2\xi\sin\phi}{\sin[(N\!+\!1)\phi]+2i\xi\sin(N\phi)-\xi^2\sin[(N\!-\!1)\phi]}\right|^{2}
\]
with $\xi\!=\!\Delta/V\!=\!(V_S/\gamma)(V_S/V)\!\ll\!1$ since for broadband leads one would naturally expect the inequalities $V_S\!<\!\gamma$ and $V_S\!\lesssim\!V$ to be fulfilled. This is the resonance tunneling regime, in which the energies of the transmission maxima are \emph{approximately} given by the roots of the equation $\sin[(N+1)\phi]\!=\!0$ (corresponding to the minima of the denominator) as follows
\begin{equation}
\phi^{\,max}_{k}=\frac{\pi k}{N+1},\,\,\,k=1,2,\cdots,N,
\label{phimax}
\end{equation}
to result in the resonance transmission band of precisely $N$ energy channels for the AW of $N$ atoms in length. They are
\begin{eqnarray}
E^{\,max}_{k}\!=E_{0}-2V\!\cos\phi^{\,max}_k=E_{0}-2V\!\cos\frac{\pi k}{N+1}\,,\nonumber\\[-0.1cm]
\label{Tprismax}\\[-0.1cm]
T^{\,max}_k=T(E^{\,max}_{k})=\frac{1}{1+\xi^{2}\cos^{2}\!\phi^{\,max}_k}\,.\hskip1cm\nonumber
\end{eqnarray}
These transmission maxima channels interchange with transmission minima given approximately by the roots of the equations $\sin[(N+1)\phi]\!=\!\pm1$ (corresponding to the maxima of the denominator)
\begin{equation}
\phi^{\,min}_{k}=\frac{\pi(k+1/2)}{N+1},\,\,\,k=1,\cdots,N-1,
\label{phimin}
\end{equation}
yielding
\begin{eqnarray}
E^{\,min}_{k}\!=E_{0}-2V\!\cos\phi^{\,min}_k\!=E_{0}-2V\!\cos\frac{\pi(k+1/2)}{N+1}\,,\nonumber\\
\label{Tprismin}\\[-1.2cm]\nonumber
\end{eqnarray}
\[
T^{\,min}_k\!=\!T(E^{\,min}_{k})=\!\frac{4\xi^{2}\sin^{2}\!\phi^{\,min}_k}{\left[1-\xi^{2}\cos2\phi^{\,min}_k\right]^2\!+4\xi^2\cos^2\!\phi^{\,min}_k}.\nonumber
\]
The magnitude of $T^{\,min}_k$ is seen to increase with $\Delta$ (as long as $\xi\!=\!\Delta/V\!\!<\!1$), thereby representing the strength of the coupling of the AW to the leads. For instance, $T^{\,min}_k\!=\!4\xi^{2}/(1+\xi^2)^2$ for $E^{\,min}_{k}$ in the center of the transmission band ($E^{\,min}_{k}\!=\!E_0$, whereby $\cos\phi^{\,min}_k\!\!=\!0$), to give $T^{\,min}_k\!\!\approx\!(2\Delta/V)^2$ for $\Delta\!\ll\!V$ (weak AW-lead coupling)~and $T^{\,min}_k\!\approx\!1$ for $\Delta\!\sim\!V$ (strong AW-lead coupling).

In the case where $|\varepsilon_0/2V|\!>\!1$, the roots $\lambda_{1,2}$ of Eq.~(\ref{lambda12}) are real. Approximating them with their respective leading terms of the power series expansions in $|2V/\varepsilon_0|\!<\!1$, one has $\lambda_{1,2}\!\approx\!(\varepsilon_0\pm|\varepsilon_0|)/2$.~Then, $d_N$ in Eq.~(\ref{dN}) is estimated to go asymptotically as
\[
d_{N}\approx\varepsilon_0^N=(E_0-E)^N.
\]
The transmission coefficient (\ref{TEpristine}) takes a non-resonant form then that scales with $N$ exponentially,
\begin{equation}
T(E)\approx4\xi^2\!\left(\frac{\varepsilon_0}{V}\right)^{\!\!-2N}\!\!\!\!\!=4\xi^2\!\left(\!\frac{E_0-E}{V}\right)^{\!\!-2N},
\label{Tprisout}
\end{equation}
showing a fast exponential decrease as $N$ increases. For $|\varepsilon_0/2V|\gtrsim\!1$, on the other hand, $\lambda_{1,2}\!=\!(\varepsilon_0\pm|\varepsilon_0|\epsilon)/2$ with $\epsilon\!=\!\sqrt{1\!-(2V/\varepsilon_0)^2}$ now being a small positive parameter, to result in the leading term
\[
d_{N}\approx(N+1)\left(\frac{\varepsilon_0}{2}\right)^N\!\!\!=(N+1)\left(\!\frac{E_0-E}{2}\right)^{\!N}
\]
of the power series expansion in $\epsilon$.~In this regime, the transmission coefficient (\ref{TEpristine}) is an energy independent constant decreasing with $N$ as follows
\[
T(E)\approx\frac{4\xi^2}{(N+1)^2}\,.
\]
This can also be obtained using Eq.~(\ref{dNpristine}) for $|\varepsilon_0/2V|\lesssim1$.

\subsection {Coupled CN--AW system \label{CNAW}}

Non-zero $\mu$ changes drastically the electron transport through the coupled CN--AW system. Intuitively, one would expect additional transmission resonances (Fano-like~\cite{Fano}) to appear in the transmission coefficient~(\ref{T}). In this section we analyze Eq.~(\ref{TEfin1}) qualitatively to show that this is indeed the case. This analysis is continued in Sections~\ref{sec5} and \ref{sec6} to discuss the numerical results.

Dividing the numerator and denominator of Eq.~(\ref{TEfin1}) by $V^N$, one obtains
\begin{equation}
T(E)\!=\!\left|\frac{2\xi\,\rho_{N-1}^{}}{\xi^2\delta_{N-2}+2i\xi\,\delta_{N-1}-\delta_N}\right|^{2},
\label{TEfin2}
\end{equation}
where $\rho_N^{}\!=S_N/V^N$ and $\delta_N\!=D_N/V^N$. With $\xi\!\ll1$, the transmission maxima are determined by the condition $\delta_N\!=0$, to minimize the denominator. If $\mu\!=\!0$, this becomes $d_N\!=0$, according to Eq.~(\ref{DN}), to bring us back to Eqs.~(\ref{Tprismax}) and (\ref{Tprisout}) for $|\varepsilon_0/2V|\!<\!1$ (resonance transmission band) and for $|\varepsilon_0/2V|\!>\!1$ (exponentially small transmission domain), respectively. For $\mu\!\ne\!0$ and $|\varepsilon_0/2V|\!>\!1$, we see from Eq.~(\ref{DN}) that there exists one more possibility to make $\delta_N$ close to zero. This is where $\varepsilon_p(\varepsilon_0+2V)=N\mu^2$, to result in two additional energy levels as follows
\begin{equation}
E_{1,2}=\frac{1}{2}\left[E_0+2V\!+E_p\pm\sqrt{\left(E_0+2V\!-E_p\right)^2\!+4N\mu^2}\,\right]\!\!.
\label{E12}
\end{equation}

As the top and bottom edges of the pristine AW tunneling band are given by $E\!=\!E_0\pm2V$ [see Eq.~(\ref{Tprismax})], the $E_1$ (higher energy) level falls into the domain $E\!>\!E_0+2V$ (or $\varepsilon_0/2V\!<\!-1$) of the exponentially small transmission of the pristine AW, thereby \emph{opening} an extra transmission channel in this opaque area. At fixed $\mu\!\ne\!0$, raising in energy with $N$, this channel stays within the CN forbidden gap as long as $N\mu^2\!<(E_g-E_p)(E_g-E_0-2V)$, crossing into the CN conduction band when the inequality changes its sign. The $E_2$ (lower energy) level falls into the resonance tunneling band $E_0-2V\!<\!E\!<\!E_0+2V$ (or, equivalently, $|\varepsilon_0/2V|\!<\!1$) of the pristine AW and remains there as long as the inequality $N\mu^2\!<4V(E_p-E_0+2V)$ holds true, lowering in energy with $N$. For large enough $N$ this inequality changes the sign, while the channel goes into the exponentially small transmission domain $E\!<\!E_0-2V$ (or $\varepsilon_0/2V\!>\!1$) of the pristine AW.

Inside the pristine AW transmission band, close to the center of the band where $|\varepsilon_0/2V|\!\ll\!1$, in Eq.~(\ref{dNpristine}) one has $\sin\phi\!=\!\sin[\arccos(\varepsilon_0/2V)]\!\approx\!\sin(\pi/2)\!=\!1$ to within terms of the second order in $|\varepsilon_0/2V|$. Then, Eq.~(\ref{dNpristine}) becomes
\begin{equation}
\frac{d_N}{V^N}\approx\sin\!\!\left[\frac{(N\!+\!1)\pi}{2}\right]\!=\cos\!\left(\!\frac{N\pi}{2}\right)\!=i^N\frac{1+(-1)^N}{2}.\!\!
\label{dNVN}
\end{equation}
Using this in Eqs.~(\ref{DN}) and (\ref{SN}) to evaluate $\rho_{N-1}^{}$, $\delta_{N}$, $\delta_{N-1}$, and $\delta_{N-2}$ in Eq.~(\ref{TEfin2}), one can simplify this equation to the form
\begin{widetext}
\vspace{-0.5cm}
\begin{equation}
T(E)\approx\frac{4\xi^2\left[\alpha_{N}^{}(E)\!+\mu^2\sin(N\pi/2)\right]^2}{\left[(1+\xi^2)q\cos(N\pi/2+\eta)+(1-\xi^2)(-1)^N\mu^2\right]^2+
4\xi^2\left[q\sin(N\pi/2+\eta)-(-1)^N\mu^2\right]^2}\,,
\label{TEapproxbandcenter}
\end{equation}
\vspace{0.15cm}
\end{widetext}
where
\begin{equation}
\alpha_{N}^{}(E)\!=\varepsilon_p(\varepsilon_0+2V)-N\mu^2=(E-E_1)(E-E_2),
\label{alphaN}
\end{equation}
$\eta=\arccos(\alpha_{N+1}^{}/q)$, and $q=\!\sqrt{\alpha_{N+1}^2\!+\mu^4}$. If $\mu\!=\!0$, then Eqs.~(\ref{dNVN})--(\ref{alphaN}) bring us back to Eqs.~(\ref{Tprismax}) and (\ref{Tprismin}) [with $\phi_k^{max(min)}\!=\pi/2$] for odd and even $N$, respectively. For non-zero $\mu$ the factor in the brackets in the numerator of Eq.~(\ref{TEapproxbandcenter}) becomes either $\alpha_{N}^{2}(E)$ if $N$ is even, or $\alpha_{N\mp1}^{2}(E)$ if $N$ is odd of the form $4n\pm1$, $n\!=\!1,2,3,...$ being positive integers. Then, in view of Eq.~(\ref{alphaN}) and the fact that the denominator of Eq.~(\ref{TEapproxbandcenter}) is always non-zero for $\mu\!\ne\!0$, the transmission coefficient $T(E\!=\!E_2)\!=\!0$~both in the former and in the latter case, for $N$ and $N\mp1$ respectively, once $N$ is fixed. Thus, the $E_2$ energy level in the pristine AW transmission band blocks the transmission entirely, resulting in the \emph{Fano resonance}, in full accord with the total resonant reflection effect of the standard Fano-Anderson model for a bound state within the continuum of scattering states~\cite{Mahan,Fano}. The Fano resonance width $\Gamma$ can be estimated from the focal parameter of the parabola one has in the numerator of Eq.~(\ref{TEapproxbandcenter}) by setting $\alpha_{N,N\pm1}^{}\!\approx\!\alpha_N^{}\!\approx\!(E_2-E_1)(E-E_2)$ for not too small $N$ in the neighborhood of $E_2$ according to Eq.~(\ref{alphaN}), whereas $\alpha_{N+1}^{}\!\approx\!\alpha_N^{}\!\approx\!0$ in the denominator. As Eq.~(\ref{TEapproxbandcenter}) is only valid in the neighborhood of $E_0$, there should be $E_2\!\approx\!E_0$, and then $E_1\!\approx\!E_p+2V$ by Vieta's theorem, to result in
\begin{equation}
\Gamma\approx\frac{\mu^2\,\kappa(\xi,N)}{|E_0-2V-E_p|}\,,
\label{Gamma}
\end{equation}
where
\begin{eqnarray}
\kappa^2(\xi,N)=\!\left[\,\cos\!\left(\!\frac{N\pi}{2}\right)\!-(-1)^N\right]^2\hskip0.75cm\nonumber\\
+\frac{1}{4\xi^2}\left[\,(1+\xi^2)\sin\!\left(\!\frac{N\pi}{2}\right)\!-(1-\xi^2)(-1)^N\right]^2\!\!\!.\nonumber
\end{eqnarray}
We see that the Fano resonance width is directly proportional to the square of the AW--CN coupling strength and varies strongly with $N$, while also being dependent on the relative position of the CN plasmon resonance energy and the pristine AW transmission band center. For $E_p\approx\!E_0+2V$, as it shows in Fig.~\ref{fig1}~(c) in particular, Eq.~(\ref{Gamma}) results in $\Gamma\!\sim\!\mu^2/V$ in full accord with the standard Fano-Anderson model~\cite{Fano}.

Outside of the pristine AW transmission band, in the domain of the exponentially small transmission where $|\varepsilon_0/2V|\!>\!1$, Eq.~(\ref{TEfin2}) can be simplified by approximating the functions $\rho_{N-1}^{}$, $\delta_{N}$, $\delta_{N-1}$, and $\delta_{N-2}$ with their respective leading terms in $|\varepsilon_0/2V|$, while keeping in mind that $\xi\!\ll1$. This brings one to the following expression
\begin{equation}
T(E)\approx\frac{\xi^2\mu^4}{\alpha_{N}^2(E)(\varepsilon_0/2V)^2+[\alpha_{N}^{}(E)\!+\mu^2]^2\xi^2}
\label{TEapproxoutside}
\end{equation}
to allow for evaluating various asymptotic regimes of the electron transfer through the additional plasmon-assisted transmission channels. One can see, in particular, that when $E\!=\!E_{1,2}$, whereby $\alpha_{N}^{}\!=0$, Eq.~(\ref{TEapproxoutside}) yields the perfect transmission $T(E_{1,2})\!=\!1$.~In the vicinity of the (more interesting) higher energy resonance transmission channel $E\!\approx\!E_1$ (and similar for $E\!\approx\!E_2$), Eq.~(\ref{alphaN}) can be written as $\alpha_{N}^{}\!\approx(E-E_1)(E_1-E_2)\!\sim\!(E-E_1)2\mu\sqrt{N}$, while $|\varepsilon_0|\!\approx\!|E_0-E_1|\!\sim\mu\sqrt{N}$, for $N$ large enough as is seen from Eq.~(\ref{E12}). This brings the transmission coefficient~(\ref{TEapproxoutside}) to the form
\begin{equation}
T(E)\approx\frac{(\Delta/N)^2}{(E-E_1)^2+(\Delta/N)^2}\,.
\label{TEapproxoutsideres}
\end{equation}
We see the plasmon-assisted transmission energy channel to have the Lorentzian lineshape of the half-width-at-half-maximum $\Delta/N$, that is proportional to the AW--lead coupling and inversely proportional to the AW length.

For energies far from $E_{1,2}$ resonances outside of the pristine AW transmission band, the function $\alpha_{N}^{}\!$ in Eq.~(\ref{alphaN}) is non-zero, allowing for two possible plasmon-mediated electron transmission regimes.~If the AW is not too long, one can approximate $\alpha_{N}^{}\!\approx\varepsilon_p(\varepsilon_0+2V)$ in Eq.~(\ref{alphaN}). Then Eq.~(\ref{TEapproxoutside}) takes the form
\begin{equation}
T(E)\approx\left[\frac{2\Delta\,\mu^2}{(E-E_p)(E-E_0)(E-E_0-2V)}\right]^2\!,
\label{TEapproxoutsideNlow}
\end{equation}
in which $|(E-E_0)/2V|\!>\!1$ and $E\!\ne\!E_p$. In this regime, the transmission coefficient shows no wire length dependence. For long enough AWs, one has $\alpha_{N}^{}\!\approx\!-N\mu^2$, which being substituted into Eq.~(\ref{TEapproxoutside}), results in
\begin{equation}
T(E)\approx\left(\!\frac{2\Delta}{E-E_0}\right)^{\!\!2}\!\frac{1}{N^2}\,.
\label{TEapproxoutsideNhigh}
\end{equation}
This algebraic ($\sim\!N^{-2}$) transmission length dependence is much slower than the exponential transmission length dependence of Eq.~(\ref{Tprisout}) for pristine AWs. It comes from the slow plasmon-mediated transmission channel narrowing $\sim\!N^{-2}$ in Eq.~(\ref{TEapproxoutsideres}).~The inverse \emph{quadratic} length dependence in Eq.~(\ref{TEapproxoutsideNhigh}) contrasts with the inverse \emph{linear} length dependence of the phonon-mediated transmission typical of quasi-1D molecular wire systems~\cite{Petrov1}.

\section{Numerical analysis \label{sec5}}

We assume that the leads and the AW are made of the same metal, and that the AW incapsulating carbon nano\-tube is end-bonded into the leads~\cite{Leonards00,Ratner04}. (Other possibilities for CN--lead contacts can be found in Ref.~\cite{Vasily14}.) The equilibrium band lineup inside the hybrid AW--CN system is then determined by the self-consistent charge redistribution through the entire metal-(CN--AW)-metal junction~\cite{Ratner04}, to position the Fermi level~$E_F$ of the system at equilibrium in the middle of the CN forbidden gap the way it occurs for unpinned semiconducting CNs~\cite{Leonards00}. We analyze two representative semiconducting CNs to show two possibilities for relative arrangement of the CN interband plasmon resonance with respect to the encapsulated AW transmission band. They are the (11,0) CN and the (16,0) CN. With energy counted from the bottom of the CN fundamental bandgap $E_g^{(11)}\!=E_g$, by summing up the first bright exciton excitation and binding energies, $1.21$ and $0.76$~eV, as reported by Ma~\emph{et al.}~\cite{Ma06} and Capaz~\emph{et al.}~\cite{Capaz06}, respectively, one arrives at $E_g\!=\!1.97$~eV for the (11,0) CN. This makes $E_0\!=\!E_F\!=\!0.985$~eV for the AW on-site energy and the equilibrium Fermi energy of the complex hybrid system of the (11,0) CN encapsulating the AW of the same metal as that of leads. For the (16,0) CN, we evaluate $E_g\!=\!1.47$~eV numerically using the ($\mathbf{k}\cdot\mathbf{p}$)-method by Ando~\cite{Ando}, to give $E_0\!=\!E_F\!=\!0.735$~eV for the AW encapsulating hybrid (16,0) CN system. For the AW, we use Na metal parameters, with the electron effective mass $m^\ast\!=\!1.0\,m_0$~\cite{Girifalco} ($m_0$ is the free electron mass) and the lattice constant $a\!=\!4.225$~\AA~\cite{Kittel}. This yields the nearest neighbor electron hopping rate $V\!=\!\hbar^2/2m^\ast a^2\!=\!0.21$~eV. We choose the AW--lead coupling to range within $\Delta\!\sim\!0.01-0.1$~eV as other authors earlier did~\cite{Muj94b,Kirczenow00}. The AW--CN coupling $\mu$ varies broadly in accordance with Eq.~(\ref{mu}) as discussed in the introduction above.

Figure~\ref{fig1}~(a),~(b),~(c) shows for the (11,0) and (16,0) CNs the low-energy real, imaginary and real inverse conductivities, and the local photonic DOS resonances originating from them, to scale with the finite-length (100 atoms) sodium AW transmission bands. To calculate the graphs in~(a), we used the $(\textbf{k}\cdot\textbf{p})$-method by Ando with the exciton relaxation time 100~fs for both CNs (consistent with earlier estimates~\cite{Ferrari07,Perebeinos07}). Many-particle Coulomb correlations are included in these calculations by solving the Bethe-Salpeter equation in the momentum space within the screened Hartree-Fock approximation as described in Ref.~\cite{Ando}. Real conductivities consist of series of peaks ($E_{11},E_{22},...$) representing the 1st, 2nd, etc., excitons. As imaginary conductivities are linked with real ones by the Kramers-Kronig relation, the real inverse conductivities show the interband plasmon peaks $P_{11},P_{22},...$ right next to $E_{11},E_{22},...~$ (first observed in Ref.~\cite{Pichler98}; see Refs.~\cite{Bondarev09,BondarevNova11,Bondarev12,Bondarev12pss,Bondarev14,Popescu11,Woods13} for more details). The graphs in (b) show the local photonic DOS functions for the excited state non-radiative spontaneous decay of a two-level dipole emitter placed on the symmetry axis (inset) of the (11,0) and (16,0) CN. Details of these calculations and similar graphs for a variety of other geometry configurations can be found in Refs.~\cite{Popescu11,Woods13,Bondarev05,BondarevNova06,Bondarev04,Bondarev06,Bondarev15OE,Bondarev07}. Comparing (a) and (b), we see the sharp single-peak DOS resonances to come from the interband plasmons of respective CNs. These are responsible for the AW--CN near-field coupling in hybrid CN systems. The coupling is due to the virtual (vacuum-type) EM energy exchange between the AW and the CN to create and annihilate plasmons on the CN surface as described by the interaction Hamiltonian~(\ref{Hint}). Comparison with~(c), which shows transmission bands for the 100 Na atoms chain calculated per Eqs.~(\ref{TEpristine}) and (\ref{dNpristine}) with $\Delta\!=\!0.05$~eV to scale $E_g$ for the (11,0) CN (top) and (16,0) CN (bottom), indicates that the 1st interband plasmon energy $E_p\!\sim\!E_F+2V$ and can be located both outside ($E_p\!\gtrsim\!E_F+2V$) and inside ($E_p\!\lesssim\!E_F+2V$) of the free AW electron transmission band.

\begin{figure}
\epsfxsize=9.0cm\centering{\epsfbox{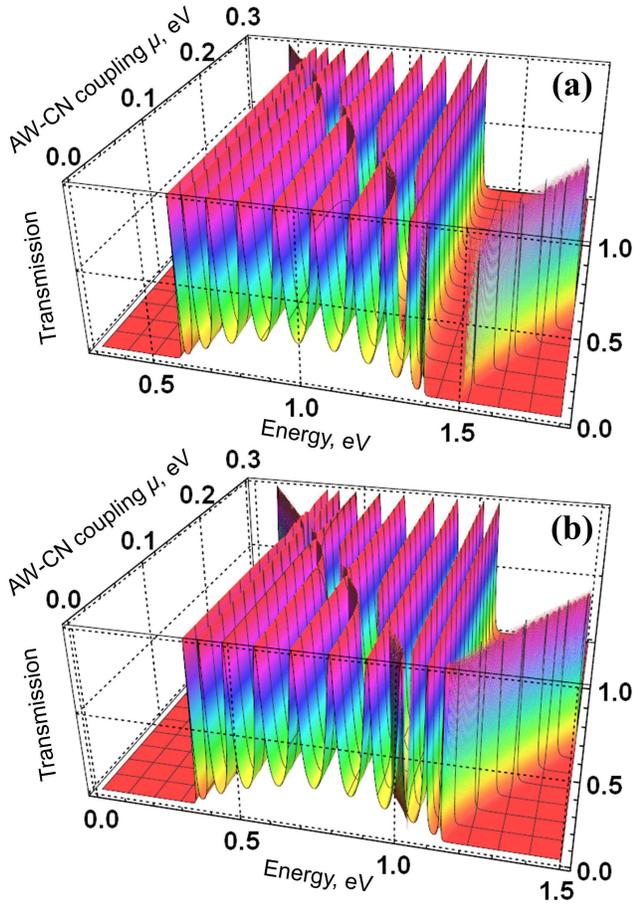}}\caption{(Color online) Transmission versus energy and AW-CN coupling strength as given by Eqs.~(\ref{TEfin1})--(\ref{lambda12}) for the AW of length $N\!=\!10$ inside the (11,0) CN [$E_p\!\gtrsim\!E_F+2V$, panel~(a)] and inside the (16,0) CN [$E_p\!\lesssim\!E_F+2V$, panel~(b)]. AW-lead coupling constant $\Delta\!=\!0.05$~eV. [Cf. Fig.~\ref{fig1}~(c)].}\label{fig2}
\end{figure}

These two possibilities are simulated and presented in Figure~\ref{fig2}~(a) and (b). Here, we show the transmission as given by Eqs.~(\ref{TEfin1})--(\ref{lambda12}) for the AW of 10 sodium atoms in length inside the (11,0) CN [(a), $E_p\!\gtrsim\!E_F+2V$] and inside the (16,0) CN [(b), $E_p\!\lesssim\!E_F+2V$] under the AW--lead coupling $\Delta\!=\!0.05$~eV with the AW--CN coupling $\mu$ varied from 0 up to 0.3~eV over the energy range to cover the entire free AW transmission band [cf. Fig.~\ref{fig1}~(c)]. At zero $\mu$, in accordance with Eqs.~(\ref{Tprismax}) and (\ref{Tprismin}), we see the free AW transmission band of 10 resonance electron transfer channels $T^{max}_{k=\overline{1,10}}\approx\!1$ separated by the transmission minima $T^{min}_{k=\overline{1,9}}$ that are controlled by the magnitude of $\Delta$ as discussed following Eq.~(\ref{Tprismin}). As $\mu$ departs from zero to increase, the near-field AW--CN interaction is seen to block some of the electron transfer channels in the AW transmission band, while opening up extra (plasmon-induced) electron transfer channels in the CN forbidden gap outside of the AW transmission band. Depending on whether $E_p$ is outside or inside of the AW transmission band, the higher energy plasmon-induced transfer channel in the CN forbidden gap either shows up gradually [panel~(a)], or splits off from the top of the free AW transmission band [panel~(b)]. The exact energies of the emerging and blocked transmission channels are given by $E_{1,2}$ in Eq.~(\ref{E12}), and their behavior is in agreement with that discussed in the previous section.

\begin{figure}
\epsfxsize=8.7cm\centering{\epsfbox{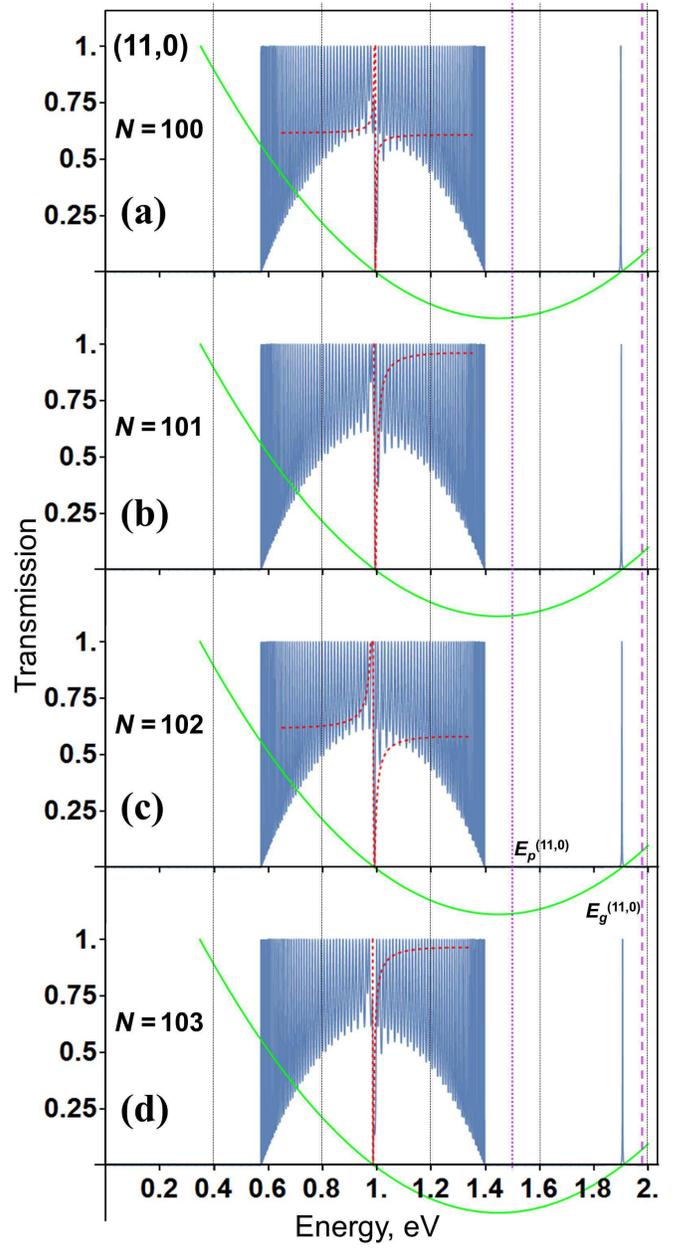}}\caption{(Color online) Transmission versus energy as given by Eqs.~(\ref{TEfin1})--(\ref{lambda12}) for the AW of varying length $N\!=\!100\!-\!103$ [(a)--(d)] inside the (11,0) CN. AW-CN and AW-lead coupling constants are $\mu\!=\!0.045$~eV and $\Delta\!=\!0.1$~eV, respectively. Red dashed lines are zone-center approximations of Eq.~(\ref{TEapproxbandcenter}). Green lines are parabolas of Eq.~(\ref{alphaN}). [Cf. Fig.~\ref{fig1}~(c)].}\label{fig3}
\end{figure}

Figure~\ref{fig3}~(a), (b), (c), and (d) illustrates in detail the Fano resonance effect discussed in the previous section. We see the transmission versus energy calculated according to Eqs.~(\ref{TEfin1})--(\ref{lambda12}) for the sodium AW of varied length $N\!=\!100$, $101$, $102$, and $103$ inside the (11,0) CN under the AW--CN coupling $\mu\!=\!0.045$~eV and the AW--lead coupling $\Delta\!=\!0.1$~eV. Dashed and dotted vertical lines trace the band gap ($E_g\!=\!1.97$~eV) and the first interband plasmon energy ($E_p\!=\!1.50$~eV) for the (11,0) CN (cf. Fig.~\ref{fig1}). Red thick dashed lines show the approximate transmission curves given by Eq.~(\ref{TEapproxbandcenter}) valid in the neighborhood of the AW transmission band center (hence the choice of $\mu$ and $\Delta$ in this calculation). Green lines are the parabolas of Eq.~(\ref{alphaN}). They are seen to intersect the abscise axis at two points, $E\!=\!E_{1,2}$ given by Eq.~(\ref{E12}). At $E\!=\!E_2$ inside the AW band, the transmission drops down to zero in view of the fact that this coupled AW--CN state (which can also be interpreted as one of the \emph{two} branches to represent the "dressed" states of the mixed CN plasmon and AW electron excitations~\cite{Bondarev06,Bondarev15OE}) is not a well defined eigen state of the entire hybrid system. An electron can occupy this state just temporarily, not permanently, as there is always a high probability for it to leave for one of the many band states that are available in this energy domain. That is why this coupled AW--CN state behaves as a scattering resonance to reflect an incident electron flux at $E\!=\!E_2$, thereby blocking transmission at this energy. Another coupled AW--CN state, the second branch of the mixed CN plasmon and AW electron excitations, is isolated at $E\!=\!E_1$ in the CN forbidden gap outside the AW band. This is a well defined eigen state of the entire hybrid system that opens up a new plasmon-mediated resonance transmission channel.

\begin{figure}
\hskip-0.37cm\epsfxsize=9.0cm\centering{\epsfbox{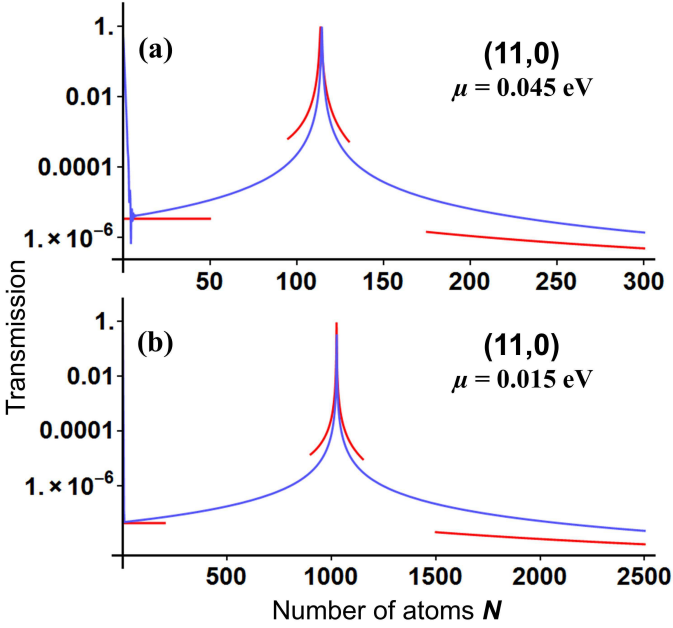}}\caption{(Color online) Log-scaled transmission versus the AW length as given by Eqs.~(\ref{TEfin1})--(\ref{lambda12}) for the AW inside the (11,0) CN at $E\!=\!1.93$~eV (CN forbidden gap outside of the AW transmission band, cf.~Fig.~\ref{fig3}).~AW--CN coupl\-ing $\mu=0.045$~eV in~(a) and 0.015~eV in (b), AW--lead coupl\-ing $\Delta=0.1$~eV. Red lines are the near-resonance and short/long-AW out-of-resonance approximations of Eqs.~(\ref{TEapproxoutsideres}) (middle line), (\ref{TEapproxoutsideNlow}) (left line) and~(\ref{TEapproxoutsideNhigh}) (right line), respectively.}\label{fig4}
\end{figure}

\begin{figure}
\hskip-0.2cm\epsfxsize=8.83cm\centering{\epsfbox{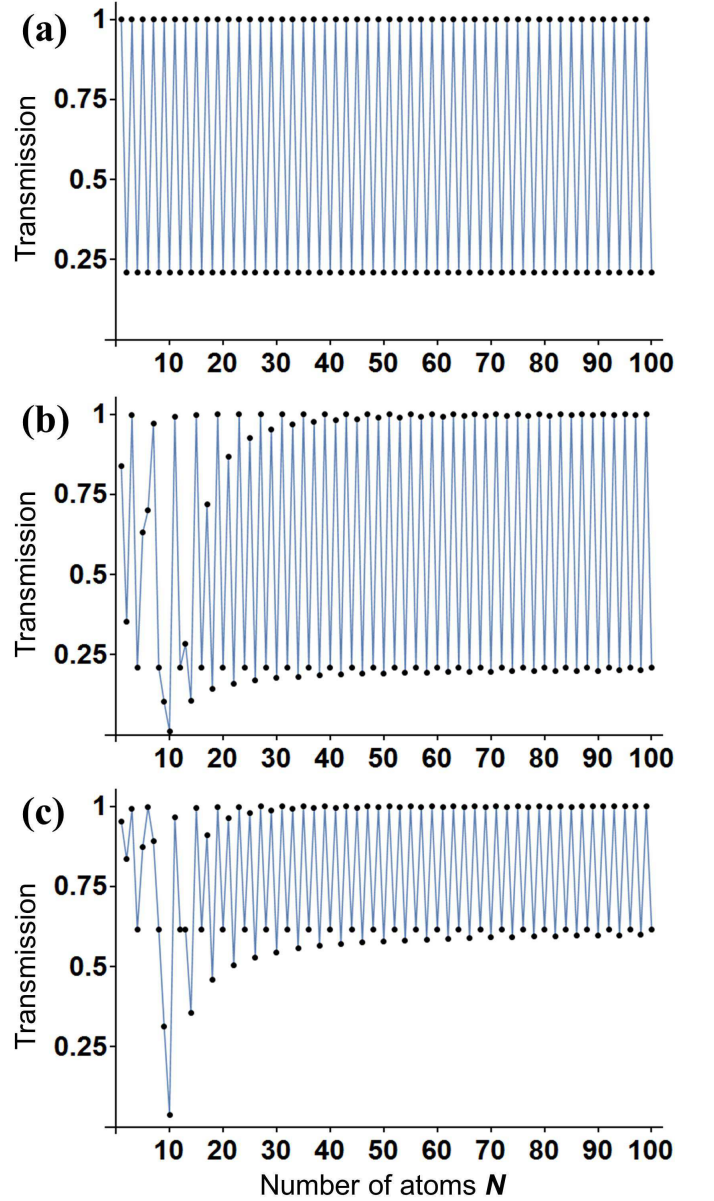}}\caption{(Color online) Transmission versus the AW length as given by Eqs.~(\ref{TEfin1})--(\ref{lambda12}) for the AW inside the (11,0) CN at $E\!=\!E_F\!=\!0.985$~eV [cf.~Fig.~\ref{fig2}~(a)]. Panel (a): $\mu\!=\!0$~eV, $\Delta\!=\!0.05$~eV. Panel (b): $\mu\!=\!0.15$~eV, $\Delta\!=\!0.05$~eV. Panel (c): $\mu\!=\!0.15$~eV, $\Delta\!=\!0.1$~eV.}\label{fig5}
\end{figure}

Figure~\ref{fig4} shows the transmission as a function of the AW length, calculated from Eqs.~(\ref{TEfin1})--(\ref{lambda12}) for the AW inside the (11,0) CN. The energy is fixed at $E\!=\!1.93$~eV, that is inside the CN forbidden gap but outside the pristine AW transmission band, and is close to the plasmon-mediated resonance transmission channel $E_1$ in Fig.~\ref{fig3}. Red lines indicate the approximations given by Eq.~(\ref{TEapproxoutsideres}) in the neighborhood of $E_1$ (middle line), and by Eqs.~(\ref{TEapproxoutsideNlow}) and (\ref{TEapproxoutsideNhigh}) away from $E_1$ in the short AW (left line) and long AW (right line) limits. In (a) the same coupling parameters as in Fig.~\ref{fig3} are used, while in (b) the AW--CN coupling is reduced by a factor of three. We see that the resonance plasmon-mediated transmission depends strongly on the AW--CN coupling strength $\mu$, and can be achieved both at shorter and at longer AW length for stronger and weaker coupling, respectively. The energy of the plasmon-mediated transmission channels is controlled by the product $N\mu^2$ as given by Eq.~(\ref{E12}). Therefore, to reach the resonance transmission regime of Eq.~(\ref{TEapproxoutsideres}) at fixed energy $E$ with $\mu$ reduced by a factor of three, one has to increase $N$ by a factor of nine. That is exactly what we see comparing (a) and (b) in Fig.~\ref{fig4}. This peculiarity is the key to practical applications of the plasmon-mediated coherent resonance transmission phenomenon.

It is interesting to see how the AW--CN coupling affects the transmission at the Fermi level energy $E\!=\!E_F$. For pristine monoatomic wires of finite length it is known, in particular, that depending on the valence and interatomic spacing their conductance shows both odd-even atom number oscillations and more complicated features such as four-atom and six-atom period oscillations (see Refs.~\cite{Datta08,Li07,Khom04,Ruitenbeek03,Lang98} and refs. therein for details]. Figure~\ref{fig5} shows the transmission coefficient versus the AW length calculated from Eqs.~(\ref{TEfin1})--(\ref{lambda12}) for the AW inside the (11,0) CN at $E\!=\!E_F\!=\!0.985$~eV. The graphs for the AW--(16,0) CN system look similar, and so are not shown here. In (a) and (b) the AW--CN coupling $\mu\!=\!0$ and 0.15~eV, respectively, while $\Delta\!=\!0.05$~eV as in Fig.~\ref{fig2}~(a). In (c) the AW--lead coupling is increased by a factor of two, $\Delta\!=\!0.1$~eV, while $\mu\!=\!0.15$~eV is the same as in (b). The odd-even atom number oscillations of the pristine AW in panel~(a) come from the oscillatory behavior of $d_N$ in Eq.~(\ref{dNpristine}) at $\varepsilon_0\!=\!0$, whereby $d_N$ is zero or non-zero to yield maximal or minimal transmission in Eq.~(\ref{TEpristine}) for odd or even $N$, respectively. In panels (b) and (c), where the AW--CN coupling is non-zero, the distinct behavior can be understood from Eq.~(\ref{TEapproxbandcenter}), which is the \emph{exact} representation of Eq.~(\ref{TEfin2}) for $E\!=E_0\!=\!E_{F}$. With $\xi^2\!\ll\!1$, this equation is seen to have maxima when
\begin{equation}
q\cos\!\left(\!\frac{N\pi}{2}+\eta\!\right)+(-1)^N\mu^2=0,
\label{cond4kplus3}
\end{equation}
where $\eta\!=\!\arccos(\alpha_{N+1}^{}/q)$ and $q=\!\sqrt{\alpha_{N+1}^2\!+\mu^4}$ with $\alpha_N^{}\!=2V(E_p-E_F)-N\mu^2$. The case where $\alpha_{N+1}^{}\!=0$~was discussed in the previous section. The transmission coefficient experiences the Fano resonance at $E\!=\!E_F$ then. This can be clearly seen in panels (b) and (c) for $N\!=\!10$ [and also in Fig.~\ref{fig2}~(a)] for the parameters chosen. For non-zero $\alpha_{N+1}^{}$ such that $\alpha_{N+1}^{2}\!\ll\mu^4$, that is for $N$ given by the inequality $|N+1-2V(E_p-E_F)/\mu^2|\ll1$, one has $q\approx\mu^2$ and $\eta\approx\pi/2$. Then Eq.~(\ref{cond4kplus3}) fulfils for all integer $n$ such that $N\!=4n+3$, yielding four-atom periodic transmission maxima one can see at low and moderate $N$ in panels (b) and (c). As $N$ increases and becomes large enough, one necessarily obtains $\alpha_{N+1}^{2}\approx\alpha_{N}^{2}\gg\mu^4$, to yield $q\approx|\alpha_N^{}|\approx N\mu^2$ and $\eta\approx\pi$. As this takes place, Eq.~(\ref{cond4kplus3}) takes the form $\cos(N\pi/2)=(-1)^N\!/N\approx0$ while the $\mu$ dependence in Eq.~(\ref{TEapproxbandcenter}) cancels, resulting in odd-even atom number transmission oscillations that are only dependent on the AW--lead coupling $\Delta$ as one can see from the graphs in panels (b) and (c).

\section {Discussion}\label{sec6}

In this study, we consider coherent electron transport through the one-atom-thick, finite-length metallic wire encapsulated in a semiconducting CN whose forbidden gap is broader than the conduction band of the wire. We ignore incoherent electron scattering processes such as those that are typical and normally studied for complex \emph{molecular} junction systems, including vibronic coupling~\cite{Petrov1,Cuniberti07}, coupling to defects~\cite{Cuniberti10}, particularities of the coupling to the leads~\cite{Kirczenow00,Cuniberti02}, etc.~\cite{Yoshizawa08} For our hybrid AW--CN system, incoherent processes like this also include electron exchange between the wire and the nano\-tube, which we do not expect to be significant due to rather strong ionization potentials of atomic metals $\sim\!5$~eV and their moderate electronegativity relative to carbon ($\sim\!0.3-1.8\!<2$)~\cite{CRChandbook}, that is insufficient to pull electrons of metal over to carbon. Overall, incoherent effects can be quite generally accounted for in our model by introducing a phenomenological finite plasmon lifetime. This redefines $E_p$ to $E_p-i\Delta E_p$ with the imaginary part representing the half-width (inverse lifetime) of the plasmon resonance, which is equivalent to the replacement
\[
\delta(\omega-E_p/\hbar)\longrightarrow\frac{1}{\pi}\frac{\Delta E_p/\hbar}{(\omega-E_p/\hbar)^2+(\Delta E_p/\hbar)^2}
\]
in Eq.~(\ref{corresp}) above. Substitution of thus modified Eq.~(\ref{corresp}) into the CN Hamiltonian~(\ref{HCNfin}) does not make any change to it provided that $\Delta E_p\ll\!E_p\,$, in which case the plasmon resonance is sharp and plasmons are well-defined long-lived excitations of the nanotube, whereby
\begin{eqnarray}
\int_0^\infty\!\!\!\!\!\frac{d\omega\,\hbar\omega\,\Delta E_p/(\pi\hbar)}{(\omega-E_p/\hbar)^2+(\Delta E_p/\hbar)^2}\nonumber\hskip0.3cm\\[0.2cm]
\approx E_p\int_0^\infty\!\!\!\!\!\frac{d\omega\,\Delta E_p/(\pi\hbar)}{(\omega-E_p/\hbar)^2+(\Delta E_p/\hbar)^2}\nonumber\\[0.2cm]
=\frac{E_p}{\pi}\left[\arctan\!\left(\frac{E_p}{\Delta E_p}\right)+\frac{\pi}{2}\right]\approx E_p,\nonumber\hskip-0.1cm
\end{eqnarray}
thus leaving our results unchanged up to terms of the first non-vanishing order in $\Delta E_p/E_p$.

In our approach, the AW is treated within the single-hopping-parameter (or single-band) tight-binding model. Such a model is realistic for the 1D chains of atoms with half-filled outermost $s$-shells. These include monovalent alkali metals and transition metals with filled $d$- (and $f$-) shells such as copper, silver, and gold. Other transition metals would feature multi-band (multi-channel~\cite{Ruitenbeek03}) conductance due to their under-filled $d$- (and $f$-) shells. However, the AW--CN near-field interaction (\ref{Hint}) is universal in its nature, and is hardly sensitive to conductance peculiarities for them to be able to affect our results.

Our main result is the prediction of the sharp Fano resonances in electron transmission through hybrid quasi-1D nanostructures of semiconducting CNs that encapsulate metal AWs. The resonances are due to the AW--CN near-field interaction in Eqs.~(\ref{Hint}) and~(\ref{mu}). The interaction couples AW electron and CN plasmon excitations to form two branches of the mixed ("dressed"~\cite{Bondarev06,Bondarev15OE}) states to represent the eigen states of the entire hybrid system. The quantity that controls the coupling is $\mu^2N$ in Eq.~(\ref{E12}). Therefore, regardless of how $E_p$ and $E_F$ are positioned relative to each other in the CN forbidden gap, a significant AW--CN coupling strength can be achieved in structures of varied length even though the single-atom coupling constant $\mu$ is small. If, for certain $\mu$ and $N$, a coupled AW--CN state falls into the AW transmission band ($E_2$ in Figs.~\ref{fig2}~and~\ref{fig3}), then, being surrounded by other band states in its vicinity, it ceases to be the well-defined eigen state; it turns into the scattering resonance to reflect an incident electron flux into the neighboring band states~\cite{Fano,Mahan}, thereby blocking the transmission at this energy. If it happens that a coupled AW--CN state is isolated outside of the AW band inside of the CN forbidden gap ($E_1$ in Figs.~\ref{fig2}~and~\ref{fig3}), then it remains a well-defined eigen state of the entire hybrid system to open up a new plasmon-mediated coherent transmission channel in the energy domain where neither of the individual pristine constituents, neither AW nor CN, is transparent. Such a resonance coherent electron transport can be quite efficient even though the actual coupling constant $\mu$, the AW length $N$, and the transmission energy are out of their resonance values since the out-of-resonance transmission coefficient falls down with $N$ relatively slowly, $\sim\!1/N^2$, as one can see from Eq.~(\ref{TEapproxoutsideNhigh}) shown in Fig.~\ref{fig4}.

The features described of the Fano resonances we predict are quite generic. They originate from the similarity between our model Hamiltonian (\ref{Htot})--(\ref{mu}) and the general Fano-Anderson model for a bound quantum state inside or outside of the continuum of scattering states~\cite{Mahan}. The Fano resonances can also manifest themselves in those metal-nanotube combinations where the AW transmission band happens to be broader than the CN forbidden gap. They may affect electron transport in the CN conduction band as well as hole transport in the CN valence band, since $E_1$ enters the CN conduction band and $E_2$ enters the CN valence band at large $\mu^2N$ (Fig.~\ref{fig2}). The result will be transmission reduction for some of the channels inside a band of states and/or an extra plasmon-mediated coherent transmission resonance in the energy domain where no band states available.

In our model, the single-atom AW--CN coupling constant $\mu$ in Eq.~(\ref{mu}) is considered to be site independent. In reality, the structure of the AWs encapsulated in the nanotube can be quite different from that of pristine AWs due to factors such as atom clustering~\cite{Belucci13}, dimerization~\cite{Vega11}, multiple atomic chains formation~\cite{Shinohara09}, as well as a variety of random spontaneous deformations of atomic chains inside the CN. In all these and other related cases, our model coupling constant $\mu$ should be considered as the effective mean interaction constant. Local deviations from the mean value due to the factors mentioned will definitely cause the inhomogeneous broadening of the Fano resonances we predict --- both inside of the AW transmission band to increase the $\Gamma$ estimate in Eq.~(\ref{Gamma}), in particular, and outside of the AW band to broaden the plasmon-mediated coherent transmission channel in the CN forbidden gap (see Figs.~\ref{fig2} and~\ref{fig3}).

Overall, by selectively controlling the AW length $N$ in the process of sample fabrication~\cite{Ruitenbeek03}, one might be able, in principle, to manipulate by the electron transport regimes as it shows in Figs.~\ref{fig4} and~\ref{fig5} and is commented above --- both inside and outside of the CN forbidden gap, both to reduce and to enhance the transmission of the hybrid AW--CN system. Controlling the AW length can also be supplemented with other external means, such as the AW transmission band tune-up through chemical or electrostatic gate control~\cite{Mujica03,Datta04}, electrostatic doping to adjust the CN forbidden gap~\cite{Spataru10}, and the quantum confined Stark effect to tune the CN plasmon energy~\cite{Bondarev12}, thus allowing for flexible transport optimization in hybrid metal-semiconductor CN systems in ways desired for practical applications.

\section{Conclusions \label{sec7}}

We study coherent electron transport through the one-atom-thick, finite-length metallic wire encapsulated into a semiconducting carbon nanotube with the forbidden gap broader than the AW conduction band.~We use matrix Green's functions formalism to develop the electron transfer theory for such a hybrid metal-semiconductor system. Our goal is to understand the inter-play between the intrinsic 1D conductance of the atomic wire and nano\-tube mediated near-field effects.

The theory we developed predicts the Fano resonances in electron transmission through the system. That is the AW--CN near-field interaction blocks some of the pristine AW transmission band channels to open up new coherent channels in the CN forbidden gap outside the AW transmission band. This makes the entire hybrid system transparent in the energy domain where neither AW nor CN is individually transparent. These generic features of the Fano resonances we predict may also manifest themselves in those metal-nanotube combinations where the AW transmission band is broader than the CN forbidden gap. They may affect both electron transport in the CN conduction band and hole transport in the CN valence band to block some of the transmission channels inside and/or to provide extra plasmon-mediated coherent transmission channels outside of bands of states. This effect can be used to control and optimize charge transfer in hybrid metal-semiconductor CN based devices for nanoscale energy conversion, separation, and storage.

\section{Acknowledgments}

M.F.G. is supported by Deutsche Forschungsgemein\-schaft through the Cluster of Excellence "Munich-Centre for Advanced Photonics" (www.munich-photonics.de). M.F.G. acknowledges the hospitality of the Department of Mathematics and Physics at North Carolina Central University, USA, where this work was started during the visit sponsored by the US NSF (ECCS-1306871). I.V.B. is supported by DOE (DE-SC0007117).

\appendix

\section {Derivation of equations~(\ref{DN}) and (\ref{SN}) \label{ApA} }

Using the matrix $\mathbf{H}$ in Eq.~(\ref{Htotmatrix}), one can derive the recursion relations for the quantities $D_{N}$ and $S_{N}$ to determine the transmission coefficient formulas~(\ref{TEfin1}) and (\ref{TEfin2}).

~

\emph{Recursion relation for $D_N$}.

Expanding the determinant $D_{N}\!=\det(\mathbf{H}\!-E)$ along~the first row, one has the set of recursion relations as follows
\begin{eqnarray}
D_{N}&=&\varepsilon_0 D_{N-1}-V A_{N} + (-1)^N\mu\,F_{N},\label{DNrr}\\[0.1cm]
A_{N}&=&V D_{N-2}-(-1)^N\mu\,F_{N-1},\label{ANrr}\\[0.1cm]
F_{N}&=&V F_{N-1} - (-1)^N\mu\,d_{N-1},\label{FNrr}\\[0.1cm]
d_{N}&=&\varepsilon_0\,d_{N-1} - V^2 d_{N-2},\label{dNrr}
\end{eqnarray}
where $A_{N}$, $F_{N}$, and $ d_{N}$ are the determinants of the $N\!\times\!N$ matrixes
\[
\mathbf{A}=\left[\begin{array}{ccccccc}
V & V & 0 & \ldots & 0 & 0 & \mu\\
0 & \varepsilon_{0} & V & \ldots & 0 & 0 & \mu\\
0 &  V & \varepsilon_{0} & \ldots & 0 & 0 & \mu\\
\vdots & \vdots & \vdots & \ddots & \vdots & \vdots & \vdots\\
0 & 0 & 0 & \ldots & \varepsilon_{0} & V & \mu\\
0 & 0 & 0 & \ldots & V & \varepsilon_{0} & \mu\\
\mu & \mu & \mu & \ldots & \mu & \mu & \varepsilon_p
\end{array}\right],
\]
\[
\mathbf{F}=\left[\begin{array}{ccccccc}
V & \varepsilon_{0} & V & 0 & \ldots & 0 & 0\\
0 & V & \varepsilon_{0} & V & \ldots & 0 & 0\\
0 & 0 & V & \varepsilon_{0} & \ldots & 0 & 0\\
\vdots & \vdots & \vdots & \ddots & \ddots & \vdots & \vdots\\
0 & 0 & \ldots & 0 & V & \varepsilon_{0} & V\\
0 & 0 & \ldots & 0 & 0 & V & \varepsilon_{0}\\
\mu &  \mu & \ldots & \mu & \mu & \mu & \mu
\end{array}\right],
\]
and
\[
\mathbf{d}=\left[\begin{array}{ccccccc}
\varepsilon_{0} & V & 0 & \ldots & 0 & 0 & 0\\
V & \varepsilon_{0} & V & \ldots & 0 & 0 & 0\\
0 & V & \varepsilon_{0} & \ldots & 0 & 0 & 0\\
\vdots & \vdots & \vdots & \ddots & \vdots & \vdots & \vdots\\
0 & 0 & 0 & \ldots & \varepsilon_{0} & V & 0\\
0 & 0 & 0 & \ldots & V & \varepsilon_{0} & V\\
0 & 0 & 0 & \ldots & 0 & V & \varepsilon_{0}
\end{array}\right],
\]
respectively. Using Eqs.~(\ref{ANrr}) and (\ref{FNrr}) to eliminate $A_N$ in Eq.~(\ref{DNrr}) results in
\begin{eqnarray}
D_{N+2}\!- \varepsilon_0 D_{N+1}\!+ V^2 D_{N}\nonumber\hskip3.17cm\\[0.1cm]
= 2(-1)^N\!\mu\,V F_{N+1}\!-\mu^2 d_{N+1},\label{DNrrfin}
\end{eqnarray}
This is the recursion relation for $D_N$. It should be solved together with recursion relations (\ref{FNrr}) and (\ref{dNrr}) under the initial conditions as follows
\begin{eqnarray}
D_0 = \varepsilon_p, \;\;\;  D_1 = \varepsilon_0 \varepsilon_p-\mu^2;\nonumber\hskip0.75cm\\[0.1cm]
F_0 = 0, \;\;\; F_1 = \mu; \;\;\;\;\;\; d_0 = 1, \;\;\; d_1=\varepsilon_0.
\label{DNrrincond}
\end{eqnarray}

~

\emph{Recursion relation for $S_N$}.

According to Eqs.~(\ref{G1Ndef})--(\ref{hN1}), quantity $S_{N-1}$ of interest is the $N1$ minor of the matrix $\mathbf{H}\!-E$. This is given by the determinant of the $N\!\times\!N$ matrix
\[
\mathbf{S}=\left[\begin{array}{ccccccc}
V & 0 & 0 & 0 & \ldots & 0 & \mu\\
\varepsilon_{0} & V & 0 & 0 & \ldots & 0 &\mu\\
V & \varepsilon_{0} & V & 0 & \ldots & 0 & \mu \\
\vdots & \ddots & \ddots & \ddots & \vdots & \vdots & \vdots \\
0 & \ldots & V & \varepsilon_{0} & V & 0 & \mu\\
0 & \ldots & 0 & V & \varepsilon_{0} & V & \mu\\
\mu & \ldots & \mu & \mu & \mu & \mu & \varepsilon_p
\end{array}\right].
\]
Expanding $S_{N-1}\!=\det(\mathbf{S})$ along the first row, one has
\begin{eqnarray}
S_{N-1} &=& V S_{N-2} - (-1)^N\mu\,B_{N-2},\label{SNrr}\\[0.1cm]
B_{N-1} &=& \mu\,d_{N-2} - V B_{N-2},\label{BNrr}
\end{eqnarray}
where $B_{N-1}$ is the determinant of the $N\!\times\!N$ matrix
\[
\mathbf{B}=\left[\begin{array}{ccccccc}
\varepsilon_{0} & V & 0 & \ldots &0 & 0 & 0\\
V & \varepsilon_{0} & V & \ldots & 0 & 0 & 0\\
0 & V & \varepsilon_{0} & \ldots & 0 & 0 & 0\\
\vdots & \vdots & \vdots & \ddots & \vdots & \vdots & \vdots\\
0 & 0 & 0 & \ldots & \varepsilon_{0} & V & 0\\
0 & 0 & 0 & \ldots & V & \varepsilon_{0} & V\\
\mu & \mu & \mu & \ldots & \mu & \mu & \mu
\end{array}\right].
\]
Combining Eqs.~(\ref{SNrr}) and (\ref{BNrr}), one arrives at the recursion relation as follows
\begin{equation}
S_{N+2} - 2VS_{N+1} + V^2 S_{N} = (-1)^N\!\mu^2 d_{N},
\label{SNrrfin}
\end{equation}
to be solved under the initial conditions
\begin{eqnarray}
S_0 = \varepsilon_p, \;\;\; S_1 = V\varepsilon_p - \mu^2;\nonumber\hskip-0.2cm\\[0.1cm]
d_0=\varepsilon_0 , \;\;\; d_1 = \varepsilon_0^2 - V^2.
\label{SNrrincond}
\end{eqnarray}
The $d_N$ initial condition is now one element downshifted [cf. Eq.~(\ref{DNrrincond})] to reflect the fact of the dimensionality reduction in Eq.~(\ref{SNrrfin}) compared to Eq.~(\ref{DNrrfin}).

~

\emph{Solving recursion relations (\ref{DNrrfin}) and (\ref{SNrrfin})}.

Recursion relations (\ref{DNrrfin}) and (\ref{SNrrfin}) are a convenient set of recursion formulas for numerical evaluation of the transmission coefficient in Eq.~(\ref{TEfin1}). They do allow for exact solution, and so they will be solved here analytically. According to Ref.~\cite{EqWorld}, the solution to the second order constant coefficient inhomogeneous recursive relation
\begin{equation}
y_{N+2}^{}\! + a\,y_{N+1}^{}\! + b\,y_{N}^{} = f_{N}^{}
\label{yrr}
\end{equation}
($a$ and $b$ are constant coefficients, $f_{N}$ is a known function) is given by the expression as follows
\begin{equation}
y_{N}^{}\! = y_{1}^{}\zeta_{N-1}^{}\! - y_{0}^{}b\,\zeta_{N-2}^{} + \sum_{k=0}^{N-2}\!f_{k}^{} \zeta_{N-k-2}^{}.
\label{yrrfin}
\end{equation}
Here
\begin{equation}
\zeta_{N}^{}\!=\frac{\lambda_1^{N+1}-\lambda_2^{N+1}}{\lambda_1-\lambda_2}
\label{zetaN}
\end{equation}
with $\lambda_{1,2}$ being the roots of the characteristic equation $\lambda^2+a\,\lambda + b = 0$. For $\lambda_1 = \lambda_2$,  Eq. (\ref{zetaN}) takes the form
\begin{equation}
\zeta_{N}^{}\!=(N+1)\,\lambda_1^{N}.
\label{zetaNlambda12eq}
\end{equation}

Starting with Eq.~(\ref{dNrr}) and bringing it to the standard form~(\ref{yrr}), one has
\[
d_{N+2}-\varepsilon_0\,d_{N+1} + V^2 d_{N} = 0.
\]
This is to be solved with initial conditions (\ref{DNrrincond}) and (\ref{SNrrincond}) for recursion relations (\ref{DNrrfin}) and (\ref{SNrrfin}), respectively. Using Eq.~(\ref{yrr}) with $f_N\!=\!0$, Eq.~(\ref{yrrfin}) and Eq.~(\ref{zetaN}), one obtains Eq.~(\ref{dN}) under initial conditions (\ref{DNrrincond}), and
\begin{equation}
d_N=\frac{\lambda_1^{N+2}-\lambda_2^{N+2}}{\lambda_1-\lambda_2}
\label{dNsol2}
\end{equation}
under initial conditions (\ref{SNrrincond}), where $\lambda_{1,2}$ are the roots of the characteristic equation $\lambda^2\!-\varepsilon_0\lambda+V^2=0$. They are given by Eq.~(\ref{lambda12}), and are subject to Vieta's formulas whereby $\lambda_1+\lambda_2=\varepsilon_0$ and $\lambda_1\lambda_2=V^2$.

Similarly, bringing Eq.~(\ref{FNrr}) to the form (\ref{yrr}), one has
\[
F_{N+2} - V F_{N+1} = -(-1)^N\mu\,d_{N+1},
\]
which should be solved under initial conditions (\ref{DNrrincond}). Then, Eq.~(\ref{yrrfin}) with $f_N\!=\!-(-1)^N\mu\,d_{N+1}$, where $d_N$ is given by Eq.~(\ref{dN}), results in
\[
F_N=\mu\,V^{N-1}\!-\mu\!\!\!\!\!\sum_{k=0}^{N\!-2\,(\ge0)}\!\!\!(-1)^k\,\frac{\lambda_1^{k+2}\!-\lambda_2^{k+2}}{\lambda_1-\lambda_2}\,V^{N-k-2}.
\]
Here, the second term can be found by summing up two geometric series with common ratios $-\lambda_1/V$ and $-\lambda_2/V$, respectively, to result in the final expression as follows
\begin{equation}
F_{N} = \frac{\mu}{\varepsilon_0 + 2 V } \left[V^{N}\!-(-1)^{N} (d_{N} + V d_{N-1})\right]\!.
\label{FNrrfin}
\end{equation}

With $F_N$ determined by Eqs.~(\ref{FNrrfin}) the right hand side of Eq.~(\ref{DNrrfin}) becomes
\[
q_{N}^{} = \frac{\mu^2}{\varepsilon_0 + 2 V }\!\left\{2V^2\!\left[d_{N} + (-1)^{N}V^{N}\right]\!\!-\varepsilon_0 d_{N+1}\!\right\}\!,
\]
which can be further rewritten as
\begin{equation}
q_{N}^{} = \frac{\mu^2}{\varepsilon_0 + 2 V }\!\left[2(-1)^{N}V^{N+2}\!-\lambda_1^{N+2}\!-\lambda_2^{N+2}\right]
\label{qN}
\end{equation}
using Eq.~(\ref{dNrr}) followed by Eq.~(\ref{dN}) to express $d_{N+2}$ and $d_N$ in terms of $\lambda_{1,2}$. With $f_N\!=\!q_N$ of Eq.~(\ref{qN}) and $\zeta_N^{}\!=\!d_N$, Eqs.~(\ref{yrr})--(\ref{zetaN}) under initial conditions~(\ref{DNrrincond}) result in the solution to Eq.~(\ref{DNrrfin}) as follows
\begin{eqnarray}
D_{N} = \left(\varepsilon_0 \varepsilon_p-\mu^2\right)d_{N-1}\! - \varepsilon_p V^2d_{N-2}\nonumber\\[0.1cm]
+ \!\!\!\!\!\sum_{k=0}^{N\!-2\,(\ge0)}\!\!\!\!\!q_{k}^{}d_{N-k-2}.\label{DNsolfin}\hskip1.5cm
\end{eqnarray}
Here, the first two terms can be written as $\varepsilon_pd_N\!-\mu^2d_{N-1}$ in view of Eq.~(\ref{dNrr}). The third term can be evaluated~by summing up the geometric series in the same way as it was done to derive Eq.~(\ref{FNrrfin}). There are three contributions to the total sum that originate from the three terms in Eq.~(\ref{qN}). Using Eqs.~(\ref{dN}) and (\ref{dNrr}) as well as the fact that $\lambda_1+\lambda_2=\varepsilon_0$ and $\lambda_1\lambda_2=V^2$, one has
\begin{eqnarray}
\sum_{k=0}^{N\!-2\,(\ge0)}\!\!\!\!\!(-1)^{k}V^{k+2}d_{N-k-2}=Vd_{N-1}\nonumber\hskip1.5cm\\
+\frac{V}{\varepsilon_0 + 2 V }\left[(-1)^NV^N-Vd_{N-1}-d_N\right]\nonumber
\end{eqnarray}
and
\begin{eqnarray}
\sum_{k=0}^{N\!-2\,(\ge0)}\!\!\!\!\!\lambda_1^{k+2}d_{N-k-2}+\!\!\!\!\!\sum_{k=0}^{N\!-2\,(\ge0)}\!\!\!\!\!\lambda_2^{k+2}d_{N-k-2}\nonumber\\[0.2cm]
=Nd_N-\varepsilon_0d_{N-1},\nonumber\hskip1.5cm
\end{eqnarray}
to result in
\begin{eqnarray}
\sum_{k=0}^{N\!-2\,(\ge0)}\!\!\!\!\!q_{k}^{}d_{N-k-2}=\mu^2d_{N-1}\,\nonumber\hskip2.5cm\\[0.1cm]
+\frac{\mu^2}{\varepsilon_0+2V}\!\left\{\!-N d_N\!+\!\frac{2V}{\varepsilon_0+2V}\!\left[(-1)^N V^N\!\!-Vd_{N-1}\!\!-d_N\right]\!\right\}\nonumber
\end{eqnarray}
after elementary algebraic simplifications. Substituting this into the right hand side of Eq.~(\ref{DNsolfin}), one finally arrives at Eq.~(\ref{DN}).

Equation~(\ref{SNrrfin}) must be solved with $d_N$ of Eq.~(\ref{dNsol2}) consistent with the initial conditions (\ref{SNrrincond}) as opposed to Eq.~(\ref{DNrrfin}) where $d_N$ on the right is given by Eq.~(\ref{dN}). Following Eqs.~(\ref{yrr}) and (\ref{yrrfin}) with $\zeta_N^{}\!=\!(N+1)V^N$, one then has the solution of the form
\begin{eqnarray}
S_{N} = \varepsilon_pV^N\!-\mu^2NV^{N-1}\hskip1.7cm\label{SNsol}\\[0.1cm]
+\,\mu^2\!\!\!\!\!\sum_{k=0}^{N\!-2\,(\ge0)}\!\!\!\!\!(-1)^k\,\frac{\lambda_1^{k+2}\!-\lambda_2^{k+2}}{\lambda_1-\lambda_2}\,(N\!-k\!-1)V^{N-k-2}.\nonumber
\end{eqnarray}
Here, the last term can be written as
\[
\frac{\partial}{\partial V}\sum_{k=0}^{N\!-2\,(\ge0)}\!\!\!\!\!(-1)^k\,\frac{\lambda_1^{k+2}\!-\lambda_2^{k+2}}{\lambda_1-\lambda_2}\,V^{N-k-1}\!,
\]
whereupon summing up the geometric series and differentiation followed by the algebraic simplifications subject to $\lambda_1+\lambda_2\!=\!\varepsilon_0$ and $\lambda_1\lambda_2\!=\!V^2$, result in the expression as follows
\[
\frac{\lambda_1\!\!-\lambda_2}{\varepsilon_0+2V}\!\left[(N\!\!-\!1)V^N\!\!+\varepsilon_0NV^{N-1}\!\!+(-1)^N\frac{\lambda_1^{N+1}\!\!-\lambda_1^{N+1}}{\lambda_1\!\!-\lambda_2}\right]\!.
\]
Substituting this into the right hand side of Eq.~(\ref{SNsol}), after simplifications one finally arrives at Eq.~(\ref{SN}).

\end{document}